\RequirePackage[2020-02-02]{latexrelease}
\documentclass[%
 aps,
 sd,%
 amsmath,amssymb,
 reprint,%
]{revtex4-1}
\usepackage{graphicx}% Include figure files
\usepackage{dcolumn}% Align table columns on decimal point
\usepackage{bm}% bold math
\usepackage{color}
\usepackage{array}
\usepackage{tabu}
\usepackage{hhline}
\usepackage{soul}
\setstcolor{red}
\usepackage{caption}
\usepackage{subcaption}
\usepackage[version=4]{mhchem}
\usepackage{braket}
\usepackage{float}

\usepackage{caption}
\newcommand{\svl}{\langle \sigma_{\rm{loss}} \ v \rangle}

\newcommand{\svlu}{\langle  \sigma_{\rm{loss}} (U)\ v \rangle}

\newcommand{\svt}{\langle \sigma_{\rm{tot}} \, v \rangle}

\newcommand{\svtrb}{\langle \sigma_{\rm{tot}} \, v \rangle_{\rm{Rb+H}_2}}

\newcommand{\svtli}{\langle \sigma_{\rm{tot}} \, v \rangle_{\rm{Li+H}_2}}

\newcommand{\grb}{\Gamma_{\rm{Rb+H}_2}}
\newcommand{\gli}{\Gamma_{\rm{Li+H}_2}}

\newcommand{\grbz}{\Gamma_{\rm{(Rb,0)}}}
\newcommand{\gliz}{\Gamma_{\rm{(Li,0)}}}

\newcommand{\svloss}{\left<\sigma_{\rm{loss}}\, v\right>}
\newcommand{\svtot}{\left<\sigma_{\rm{tot}}\, v\right>}

\newcommand{\gloss}{\Gamma_{\rm{loss}} }
\newcommand{\gtot}{\Gamma_{\rm{tot}} }
\newcommand{\gexp}{\Gamma_{\rm{exp}} }

\newcommand{\PQDU}{p_{\rm{QDU6}}}

\newcommand{\uud}{U/U_{\rm{d}}}
\newcommand{\ud}{U_{\rm{d}}}

\newcommand{\ig}{\rm{i_{\rm{g}}}}

\newcommand{\Ng}{n_{\rm{H}_2}}

\newcommand{\kb}{k_{\rm{B}}}

\newcommand{\bea}{\begin{eqnarray}}
\newcommand{\eea}{\end{eqnarray}}

\newcommand{\ma}{m_{\mathrm{t}}}
\newcommand{\mb}{m_{\mathrm{bg}}}
\newcommand{\vb}{v_{\mathrm{bg}}}
\newcommand{\nb}{n_{\mathrm{bg}}}

\newcommand{\fr}{f_{\mathrm{recap}}}

\newcommand{\sjb}{\sigma_{\mathrm{JB}}}

\newcommand{\svlE}{\left<\sigma_{\rm{loss}} (E_{\rm{max}}-E)\ v\right>}
\newcommand{\svlEt}{\left<\sigma_{\rm{loss}} (E_{\rm{max}}-E)\ v\right>}

\newcommand{\numin}{\nu_\mathrm{min}}

\newcommand{\drepump}{\delta_{\mathrm{repump}}}
\newcommand{\dpump}{\delta_{\mathrm{pump}}}

\newcommand{\vmotli}{V_{\mathrm{MOT, Li}}}
\newcommand{\vmotrb}{V_{\mathrm{MOT, Rb}}}

\newcommand{\umax}{U_{\mathrm{max}}}

\begin{document}

%\preprint{}
%\title{Cross-calibration of collision cross-sections between trapped atoms and deviation from quantum diffractive collision universality for light particles} %Title of paper
\title{Cross-calibration of atomic pressure sensors and deviation from quantum diffractive collision universality for light particles} %Title of paper


\author{Pinrui Shen$^1$, Erik Frieling$^1$, Katherine R. Herperger$^1$,  Denis Uhland$^1$, Riley A. Stewart$^1$, Avinash Deshmukh$^1$, Roman V. Krems$^2$, James L.~Booth$^3$, and Kirk W.~Madison$^1$}
%\email[Corresponding author:]{James\_Booth@bcit.ca}
%\thanks{}
\affiliation{$^1$Department of Physics and Astronomy, University of British Columbia, 6224 Agricultural Road, Vancouver, B.C., V6T 1Z1, Canada}
\affiliation{$^2$Department of Chemistry, University of British Columbia, 6224 Agricultural Road, Vancouver, B.C., V6T 1Z1, Canada}

\affiliation{$^3$Department of Physics, British Columbia Institute of Technology, 3700 Willingdon Avenue, Burnaby, B.C. V5G 3H2, Canada.}


\date{\today}

\begin{abstract}
{The total room-temperature, velocity-averaged cross section for atom-atom and atom-molecule collisions is well approximated by a universal function depending only on the magnitude of the leading order dispersion coefficient, $C_6$.  This feature of the total cross section together with the universal function for the energy distribution transferred by glancing angle collisions ($\PQDU$) can be used to empirically determine the total collision cross section and realize a self-calibrating, vacuum pressure standard.  This was previously validated for Rb+N$_2$ and Rb+Rb collisions.  However, the post-collision energy distribution is expected to deviate from $\PQDU$ in the limit of small $C_6$ and small reduced mass. Here we observe this deviation experimentally by performing a direct cross-species loss rate comparison between Rb+H$_2$ and Li+H$_2$ and using the \textit{ab initio} value of $\svtli$. We find a velocity averaged total collision cross section ratio, $R = \svtli : \svtrb = 0.83(5)$. Based on an \textit{ab initio} computation of $\svtli = 3.13(6)\times 10^{-15}$~m$^3$/s, we deduce $\svtrb = 3.8(2) \times 10^{-15}$~m$^3$/s, in agreement with a Rb+H$_2$ \textit{ab initio} value of $\langle \sigma_{\mathrm{tot}} v \rangle_{\mathrm{Rb+H_2}} = 3.57 \times 10^{-15} \mathrm{m}^3/\mathrm{s}$.
%$\svtrb = 3.60(7)\times 10^{-15}$~m$^3$/s. 
By contrast, fitting the Rb+H$_2$ loss rate as a function of trap depth to the universal function we find $\svtrb = 5.52(9) \times 10^{-15}$~m$^3$/s. Finally, this work demonstrates how to perform a cross-calibration of sensor atoms to extend and enhance the cold atom based pressure sensor.}

%The velocity averaged collision cross section, at room temperature, between neutral atoms and molecules, interacting via the Van der Waals potential, is well approximated by a universal function depending only on the magnitude of the leading order dispersion coefficient, $C_6$.  This feature of the total cross section together with the universal function for the energy distribution transferred by glancing angle collisions ($\PQDU$) can be used to empirically determine the total collision cross section and realize a self-calibrating, vacuum pressure standard.  This has been independently validated for Rb+N$_2$ and Rb+Rb collisions where the cross polarizability of the partners is large; however, the post-collision energy distribution is expected to deviate from $\PQDU$ in the limit of small $C_6$ and small reduced mass. Here we study Rb+H$_2$ collisions and observe such deviations both theoretically and experimentally.  Specifically, the total cross section for Rb+H$_2$ collisions found previously by fitting to the $\PQDU$ function is 25\% larger than that found by a cross-species sensor-atom comparison utilizing both Rb and Li sensor atoms exposed to the same background gas.  This result is timely as work is presently underway to establish accurate values for cross sections given an array of background species for specific sensor atoms, and understanding the limits of universality is essential to using it appropriately.
\end{abstract}

%\keywords{}

\maketitle


%\textcolor{magenta}{\textbf{[Note for everyone:] }My final averaged thermalized rate coefficient values are converged to 4 significant figures, and can be written as: $\langle \sigma_{\mathrm{tot}} v \rangle=$3.101$\times10^{-9}$ cm$^3$/s for Li+H2 and $\langle \sigma_{\mathrm{tot}} v \rangle=$3.570$\times10^{-9}$ cm$^3$/s for Rb+H2. \textbf{Katherine}}

\section{Introduction}

Laser cooled atoms confined in a magnetic or magneto-optical trap can be used as sensors of 
absolute pressure and particle flux measurements in vacuum \cite{Booth2011,Madison2012,Arpornthip2012,yuan2013,Rowan2015,Makhalow2016,Makhalov_2017,Julia2017,Julia2018,pub.1105615573,Booth2019,Shen_2020,Shen_2021,BARKER2021100229,zhang2022,doi:10.1116/5.0095011}. 
For example, the background gas density or pressure can be determined by observing loss of trapped atoms induced by collisions with ambient atoms or molecules.
In general, the quantitative determination of the background gas density or pressure requires the knowledge of the thermally averaged total collision cross section between the trapped atoms and all atoms or molecules in the background gas.

The total collision cross sections can be obtained from quantum scattering calculations \cite{PhysRevA.99.042704,PhysRevA.101.012702,PhysRevA.105.039903,PhysRevA.105.029902}. However, such calculations are affected by the uncertainty in the underlying potential energy surfaces (PESs) and are limited to molecular species with a small number of active degrees of freedom. As an alternative, the total collision cross sections can be obtained from measurements with a background gas of known density. However, this requires an independent calibration of gas density. 

We have recently demonstrated a different approach to determine the collision cross section by measuring the dependence of loss of trapped atoms on the trap depth. 
Such measurements rely on the universal function ($\PQDU$) for the energy distribution imparted to the sensor atom by quantum diffractive collisions mediated by van der Waals interactions \cite{Booth2019}.
 This was used to demonstrate a cold-atom based, self-calibrating, primary vacuum pressure standard \cite{Booth2019,madison2018,Shen_2020,Shen_2021}.  The validity of this pressure standard was proven for Rb+N$_2$ collisions by comparison with an orifice flow pressure standard using an ion gauge calibrated for N$_2$ using both standards.  The underlying principle, the universality of quantum diffractive collisions as manifested in the universality of the $\PQDU$ function, has also been independently validated for Rb+Rb collisions \cite{Stewart_2022}.

This collision universality relies on the collision cross section being dominated by scattering determined by the long range part of PES, parameterized by $C_6$, and on the velocity averaging of the cross-section inherent in the trap loss measurement process. The velocity averaging removes the oscillatory variation of the cross section with collision energy. These ``glory'' oscillations arise from interference between the scattering amplitudes due to the short- and long-range parts of the interaction potential \cite{Child1974}.  Therefore, deviations from the universal prediction are expected in the limit of small reduced mass and small $C_6$ \cite{Booth2019}.  
For collisions with a small reduced mass, the glory oscillation amplitude and period can become so large that the total cross section does not average to the underlying trend set by the long-range interaction.  
This effect is compounded by the relatively small atomic polarizabilities of low mass collision partners yielding relatively low $C_6$ values, enhancing the effects of the core repulsion and compromising the universal description. 

The main goal of this work is to obtain experimental evidence for the deviation of the total collision cross section for light collision partners from the values derived from the universal $\PQDU$ function. In order to do this, we require an independent measurement of the total collision cross section.
We consider Rb+H$_2$ collisions and develop an experiment to obtain the collision cross section independently of $\PQDU$ by cross-calibration with measurements for another trapped sensor atom. Specifically, we present measurements utilizing Rb and Li sensor atoms, both exposed to the same H$_2$ gas. We then use known information on Li+H$_2$ collision cross sections to obtain the total collision cross section for Rb+H$_2$  collisions. The results obtained experimentally are confirmed by rigorous quantum scattering calculations.  We show that the results thus obtained are inconsistent with the value of the cross section obtained from the $\PQDU$ function, thus demonstrating the departure of Rb+H$_2$ diffractive collisions from universality.

\section{Review of QDU} \label{QDU}

The physical system considered here involves trapped atoms (Rb and Li) as probe (sensor) particles of an ambient atomic or molecular gas at room temperature. 
We define the total collision cross section $\sigma_{\rm tot}$ as the sum of cross sections for elastic and all possible inelastic collisions between trapped sensor atoms and atoms or molecules in the background gas. 
Under certain conditions, as described in Refs.~\cite{Booth2019,Shen_2020}, the total collision rate is (universal) determined by the leading term in the long-range interaction between neutral collision partners $C_6$ (where $C_6$ is  {an} orientation angle averaged value in the case of molecules whose interactions are anisotropic) as follows:
\begin{eqnarray}
\gtot = \nb \svt
\end{eqnarray}
with
\begin{eqnarray}
\svt &\approx& 10.45812\left(\frac{C_6}{\hbar} \right)^{\frac{2}{5}} \left(\frac{\kb T}{\mb}\right)^{\frac{3}{10}} \nonumber \\
 & & + 6.71531\left(\frac{C_6}{\hbar} \right)^{\frac{1}{5}} \left(\frac{\mb}{\kb T}\right)^{\frac{1}{10}} \left(\frac{\hbar}{\mu}\right),
 \label{eq:svt_C6}
 \end{eqnarray}
where $\nb$ is the density of the background gas, $\mu$ is the reduced mass of the collision pair, $\mb$ is the mass of the background particle, and the angle brackets denote the average over the collision velocities with the Maxwell-Boltzmann distribution for room temperature  {(300 K)}. 

The measurable depletion of sensor atoms from a trap is quantified by the loss rate $\gloss$ and can be written as
\begin{eqnarray}
\gloss(U) = \nb \svlu,
\end{eqnarray}
where $\sigma_{\rm loss}$ is the total cross section for collisions that eject sensor atoms from the trap. The loss rate $\gloss$ depends on the trap depth $U$. As $U \rightarrow 0$, $\svloss \rightarrow \svtot$ and $\gloss \rightarrow \gtot$.  In this limit, every collision liberates a trapped sensor atom.

As was shown in Refs.~\cite{Booth2019,Shen_2020}, the trap depth dependence of $\gloss$ (for small values of $U$) is a universal function 
\begin{eqnarray}
\nonumber
\svlu &\approx& \svt \left[1-\sum_{j=1}^{\infty}\beta_j\left(\frac{U}{\ud}\right)^j\right],\\
 &\approx& \svt\left[1- \PQDU\left( \frac{U}{\ud} \right) \right]
\label{eq:svl_universal}
\end{eqnarray}
where the coefficients $\beta_j$ are given in Tab.~\ref{beta} and $\ud$ is the characteristic energy scale for quantum diffractive collisions,
\begin{eqnarray}
\ud &=& \frac{4\pi \hbar^2}{\ma \bar{\sigma} }\ = \  \frac{4\pi \hbar^2 v_p}{\ma \svt}.
\label{eq:Ud}
\end{eqnarray}
Here $\ma$ is the mass of the trapped atom and $v_p=\sqrt{2\kb T/\mb}$ is the most probable velocity of the background particles (of mass $\mb$ at temperature $T$).  The expansion in Eq.~\ref{eq:svl_universal} is valid for $U \le \ud$. The universal function $\PQDU$ is the cumulative energy distribution function of the trapped sensor atom after collision, representing the probability that the sensor atom remains in a trap of depth $U$ after the collision.  The trap loss probability is $(1-\PQDU)$.

Ref. \cite{Booth2019} showed that the coefficients $\beta_j$ are the same for different atomic and molecular species, 
provided the strength of the van der Waals interaction is significant.

\begin{table}[!ht]
\centering
\begin{tabular}{|c|c|}
\hline
Term & $\beta_j$  \\
\hline
1 & 0.673 (7)  \\
2 & -0.477 (3) \\
3 & 0.228 (6)\\
4 & -0.0703 (42)\\
5 & 0.0123 (14)\\
6 & -0.0009 (2)\\
\hline
\end{tabular}
\caption{The universal coefficients appearing in Eq. \ref{eq:svl_universal} were derived from fitting the quantum scattering (QS) computations for $\svlu$ to a sixth order polynomial.  The derivation of $\beta_j$ can be found in  \cite{Booth2019}.}
\label{beta}
\end{table}

This feature of the velocity averaged cross sections allows one to measure $\svt$ by recording the trap loss rate $\gloss = \nb \svlu$ for different trap depths, normalizing these measurements by the total collision rate (i.e.~the loss rate at zero trap depth, $\gtot$), and fitting the ratio $\gloss/\gtot = \svlu/\svt$ to the universal shape, Eq.~\ref{eq:svl_universal} \cite{Booth2019}.  The power of this approach is that any test gas species can be calibrated following this prescription with no need for \textit{ab initio} computations requiring \textit{a priori} knowledge of the interaction potentials. Once the $\svt$ are determined  experimentally, the loss of atoms from a magnetic trap with a fixed depth $U$ can be used to determine the gas pressure via Eq.~\ref{pressurestd},
\begin{eqnarray}
P & = & n k_{\rm{B}} T = \left(\frac{\gloss(U)}{\svlu}\right) k_{\rm{B}}T.
\label{pressurestd}
\end{eqnarray}

\subsection{Origin of the Quantum Diffractive Collision Universal Law}{\label{APQDU}}

\begin{figure}
\includegraphics[width=0.45\textwidth]{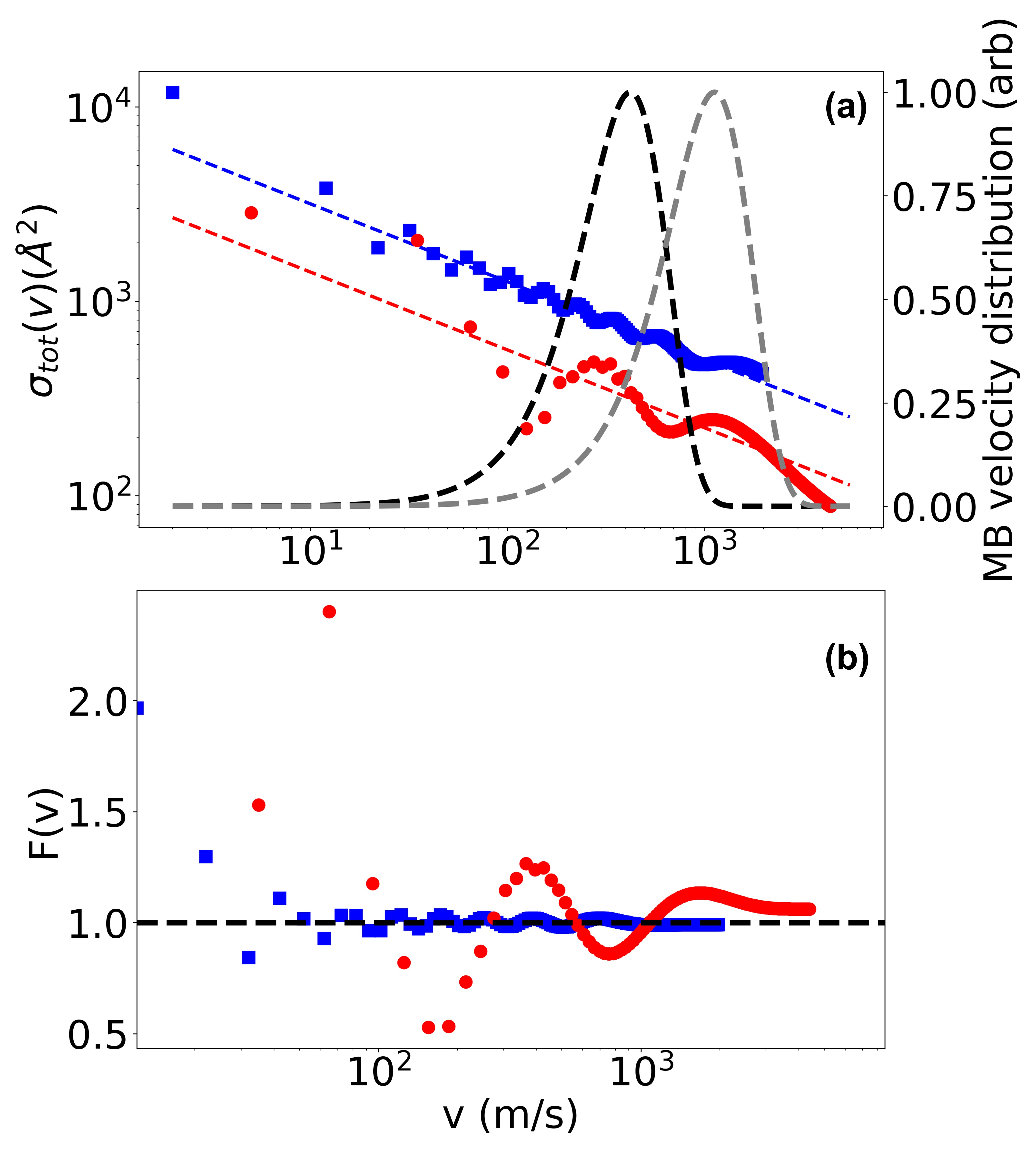}
\caption{
The total collision cross section as a function of relative velocity is shown in (a) for Rb-Ar (blue) and Rb-He (red) collisions on a log-log plot.  The cross sections exhibit oscillations around a trend line modeled by the Jeffries Born approximation $\sjb \sim \left[ \frac{C_6}{\hbar v} \right]^{2/5}$ (thin dashed lines).  Also shown are the Maxwell-Boltzmann velocity distributions for the relative collision velocity given a stationary Rb sensor atom and room temperature gas of Ar (thick black dashed line) and for He (thick grey dashed line) atoms.  The cumulative ratio $F(v)$ given by Eq.\ref{cumulativeratio} is shown in (b).  Its value at $v=\infty$ is the ratio of the true total velocity averaged cross section $\svt$ to that computed with the Jeffries Born approximation.  When the velocity averaged quantity $\svt$ is computed, for Rb-Ar collisions (blue), we see that $F(\infty) \approx 1$ meaning that it is independent of the oscillations and only depends on the average velocity and the magnitude of $C_6$.  For Rb-He, a residual dependence remains and $F(\infty)\ne1$ because the larger oscillation amplitude and period is no longer averaged away by convolution with the MB distribution.
}
\label{fig:origins}
\end{figure}

The insensitivity of the velocity-averaged total and loss cross sections to the short range details of the PES arises from velocity averaging in a regime
where the global variation of the cross section follows a trend set by the long range PES.  To illustrate this, consider Fig.~\ref{fig:origins}, which shows a log-log plot of a single-channel quantum scattering (QS) calculation of the total collision cross section  % {\textbf{[$\leftarrow$ short form QS already introduced in Table 1 caption]}} 
as a function of relative collision velocity for Rb-Ar (the solid blue traces) and for Rb-He (the solid red traces) collisions. These QS computations are performed using the Lennard-Jones shape PES for Rb-Ar and Rb-He with the long-range coefficient, $C_6$, and the potential depth information from Ref.~\cite{C8CP04397C}. These computed cross sections reveal the glory oscillations superimposed on the Jeffries-Born (JB) prediction given in Eq.~\ref{eq:sigmaJB} and shown by the dashed red and blue lines in Fig.~\ref{fig:origins}(a).
\begin{eqnarray}
\sjb &\sim & 8.0828 \left[ \frac{C_6}{\hbar v} \right]^{2/5}.
\label{eq:sigmaJB}
\end{eqnarray}
Note that at much higher velocities, the variation of the cross section with velocity follows a different %decreasing
trend set by the short range details of the PES \footnote{At relative velocities high enough that the de Broglie wavelength of the collision complex is on the order of the bound state minimum distance, the cross section follows the trend $\sigma \sim \left[ \frac{C_{12}}{\hbar v} \right]^{2/11}$ for a Leonard-Jones potential.}

%\begin{center}
%\begin{figure}[!ht]
%\end{center}

It is clear that the character of the glory oscillations is very different for these two species. 
The oscillation amplitudes and phases are determined by the relative magnitude and phase of the long and short range scattering contributions, and the number of oscillations is related to the number of rotation-less bound states supported by the potential \cite{Child1974}.  For Rb-Ar, where the $C_6$ coefficient is close to a factor of 10 larger than that for Rb-He collisions \footnote{The theoretically predicted values for C$_{6}$/$E_H a_0^6$ are 334(2) and 44.07(11) for Rb-Ar and Rb-He respectively.} the total cross-section is dominated by the long range contribution. Coupled with the the larger reduced mass of the Rb-Ar collision system one observes a larger number of lower amplitude glory oscillations, than those associated with Rb-He collisions.
 
The trap loss rate measurements are inherently averaged over the Maxwell-Boltzmann (MB) speed distribution of the room temperature background gas (the dotted traces in Figure~\ref{fig:origins}).
When this average encompasses many oscillation periods and the oscillation amplitudes are small, their effect is minimized and the information about the short range PES carried by the glory oscillations is removed.
In these cases, the quantity $\svt$ is universal in that it 
depends primarily on the average relative collision velocity and long-range character of the PES set by the magnitude of $C_6$.  By contrast, a small $C_6$ yields in larger amplitude glory oscillations, and a small reduced mass results in fewer oscillations within the MB velocity distribution. Thus, the averaged quantity $\svt$ retains information about the short range portion of the PES and will deviate from the universal prediction.

To illustrate this point, consider Fig.~\ref{fig:origins}(b).  It shows the cumulative integral ratio
\bea
%F(v) & = & \frac{\int_0^{v} 4 \pi \vb^2 \sigma(\vb) \rmb(\vb) d\vb}{\int_0^{v} 4 \pi \vb^2 \sjb(\vb) \rmb(\vb) d\vb}\\
F(v) & = & \frac{\int_0^{v} 4 \pi \vb^3 \sigma(\vb) e^{- \mb \vb^2 / 2 \kb T } d\vb}{\int_0^{v} 4 \pi \vb^3 \sjb(\vb) e^{- \mb \vb^2 / 2 \kb T } d\vb}
\label{cumulativeratio}
\eea
where $\sigma(v)$ is the QS computed value (shown in (a) by the solid lines) and $\sjb(v)$ is the Jeffries Born approximation (shown in (a) as the dashed lines) to the velocity dependent cross section.  The value of this function $F_{\infty}=F(\infty)$ denotes the ratio of the total velocity averaged cross section $\svt$ computed from the full PES using the QS calculations to that found using the approximation.  For Rb-Ar collisions, we find that $F_{\infty} \approx 1$ indicating that the total velocity averaged cross section is independent of the oscillations and thus independent of the short range PES.  For Rb-He collisions, because the oscillation amplitude and period is larger, $F(\infty)\ne1$.  This indicates that a residual dependence on the short range PES remains.

%The exact ratio of the true value for $\svt$ and the value predicted from the Jeffries Born approximation is unknown because the true PES is unknown.  However, bounds can be put on the size of the discrepancy. 

Since the velocity-averaged loss cross section asymptotically approaches the velocity-averaged total cross section as the trap depth goes to zero, $\svl$ is expected to also be insensitive to the short range details of the PES at small trap depths.  The trap-depth scale over which this insensitivity persists is set by $\ud$.  In brief, collisions that impart energies of $\ud$ or less cannot carry any information about variations of the potential on a length scale smaller than the total cross section.   {This can be understood as a consequence of the Born approximation.}

Because universality results from the velocity averaging, the parameter regime in which this universality is expected to break down is small $C_6$, small reduced mass, and narrow velocity distributions (i.e.~very cold background gases).  Although Ref.~\cite{Booth2019} explored the robustness of this universality for the case of Rb-Ar collisions by varying the PES at short range (thus changing the relative magnitudes of the short and long range scattering contributions) and by varying the test gas temperature, the regime of low reduced mass was only considered analytically and only briefly.  Therefore, this work focuses on the low reduced mass complexes Rb-H$_2$ and Li-H$_2$.  A full study of the parameter regimes where universality holds is needed; however, such a study is beyond the scope of this paper.

\section{Experimental Procedure}\label{exp}

Two experimental apparatuses were used in this work.  Device (A), described in Refs.~\cite{Booth2019,Shen_2020}, is only capable of trapping Rb atoms. It was used to perform an extensive study of the trap depth dependence of the loss rate of Rb sensor atoms exposed to a H$_2$ background gas in order to extract the value of $\svt$ based on a QDU universality description. Device (B), described in Refs.~\cite{doi:10.1063/1.4945567}, was used to perform loss rate measurements of both Rb and Li atoms exposed to the same H$_2$ background gas, thus providing a direct measurement of $\svtli : \svtrb$.  Here we give a brief description of the two devices and the experimental protocols used in this study.

\subsection{Device (A)}\label{sec:deviceA}

\begin{figure*}[!ht]
\includegraphics[width=\textwidth]{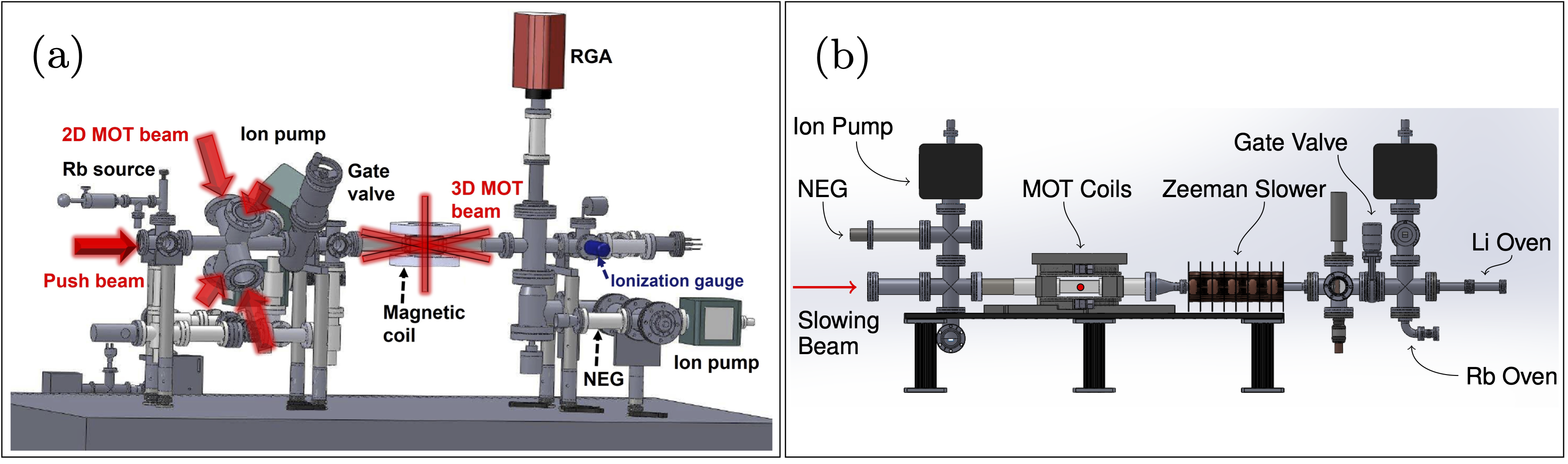}
\caption{Experiment A and B setup.  Panel (a) shows a schematic of the experiment apparatus A including the Rb source stage, the 2D MOT section, the 3D MOT section (sensing section) and the pumping section. 2D, 3D MOT light and the push light are indicated by red arrows. H$_2$ gas is introduced into the system by heating the NEG, located on the bottom side of the apparatus. The RGA installed on the top side of the apparatus is used to monitor the partial pressures of all gases introduced into the system by heating the NEG. The pressure change of the H$_2$ gas is recorded by the ionization gauge (IG).  Panel (b) shows the apparatus of device B. Rb and Li atoms leave the ovens on the right, passing through a differential pumping tube and through the Zeeman slower until being captured by the MOT.  The NEG (shown) on the left was heated to increase the H$_2$ partial pressure in the MOT region.
}
\label{experiment_setup}
\end{figure*}

Device (A) consists of an atom source (a 2D MOT) that is separated from and provides a cold flux of atoms to the pressure measurement section composed of a magneto-optic trap (a 3D MOT), a co-located magnetic trap (MT), and a suite of ionization gauges (IGs), including a residual gas analyzer (RGA), illustrated in Fig.~\ref{experiment_setup}(a). The H$_2$ background gas was introduced by applying a voltage to the non-evaporate getter pump (NEG, C400-2-DSK from SAES Getters) that heats up the NEG and releases the absorbed gas species from the getter. We used the RGA to detect which gases were present in the system when the NEG is heated. We verified that H$_2$ is the dominated gas species. For applied voltages between 2.5~V and 4.5~V, the total background pressure contributions from other gas species is less than 1~\%.

\begin{figure}[ht!]
\includegraphics[width=0.45\textwidth]{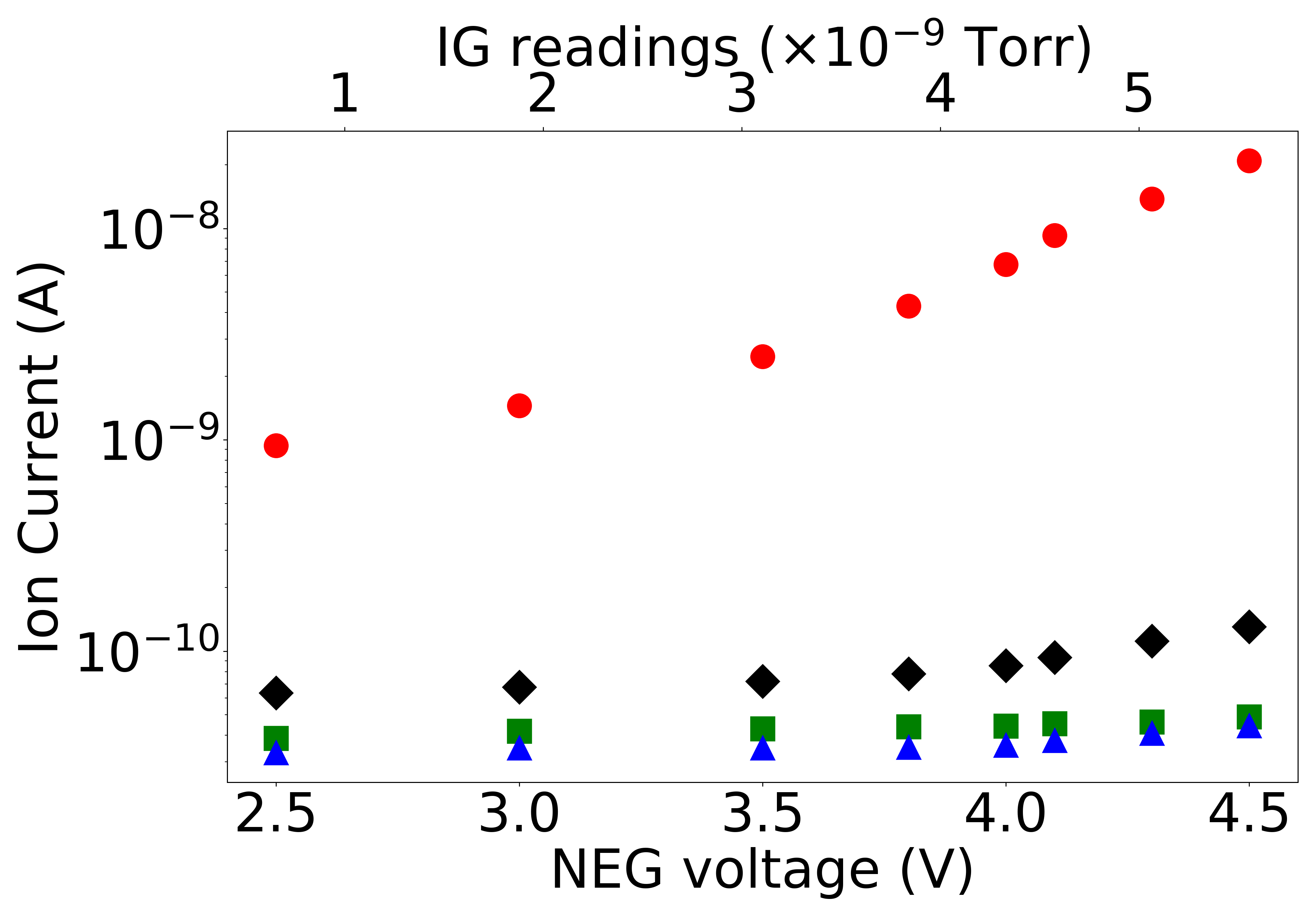}
\caption{Measured RGA signal as a function of the voltage applied across the NEG. At each applied voltage, an RGA measurement is performed to determine the partial pressure of the background gas species. At the same time, the ion gauge (IG) is used to record the total background pressure.  Red circles represent the H$_2$ gas, green squares represent the O$_2$ gas, blue triangles represent the H$_2$O gas, and the black diamonds represent the CO gas. For applied voltages in the range of [2.5 V, 4.5 V] we observe that the background pressure contribution from other gas species is less than 1~\%.  We observed that when the NEG voltage is larger than 4.5 V, the partial pressure of CO exceeds 1~\% of the total.}
\label{NEGdata}
\end{figure}

As discussed in  {Refs.}~\cite{Booth2019,Shen_2020}, to create a cold ensemble of sensor atoms, we load a 3D MOT from a cold atomic beam emerging from a 2D MOT.  The MOT trapping (or ``pump") laser beams for both the 2D and 3D MOT are locked $\delta =12$~MHz below the D$_2$ (5$^2$S$_{1/2}$ $\rightarrow$ 5$^2$P$_{3/2}$) $F=2\rightarrow F'=3$ transition of the $^{87}$Rb sensor atoms. The ``repump" laser light is tuned to resonance with the $F=1\rightarrow F'=2$ transition.  The magnetic field configuration is a spherical quadrupole for the 3D MOT and is operated with a gradient of 13.6 G/cm along the axial direction. During the measurement sequence, approximately $10^7$ atoms are loaded into the 3D MOT, producing a fluorescence signal from an amplified photo-detector of $V_{\rm{MOT}}$. Using a technique described in Ref.~\cite{Jooya2013}, we verified that our photodetector signal is linear in the atom number.  This is an essential requirement to correctly deduce the atom loss rate from the decay of the fluorescence signal.

After loading the 3D MOT, the atoms are cooled by turning down the pump laser power and changing the detuning of the pump laser frequency to $\delta=60$~MHz below the $F=2\rightarrow F'=3$ transition transition for 20 ms. The pump light is then extinguished while leaving the repump light on. At the same time, we apply a right-hand circularly polarized optical pumping (OP) beam which is tuned into resonance with the $F=2\rightarrow F'=2$ transition.  An additional magnetic field, generated by a Helmholtz coil pair is also introduced along the propagation direction of the OP beam in order to drive $\sigma_+$ transitions.  After 2 ms, the repump light, the optical pumping light and the Helmholtz coils are extinguished and the quadrupole magnetic field gradient is increased.  Atoms that fall into the low-field seeking sublevels ($m_F=1$ and $m_F=2$) of the $F=2$ manifold are retained in the  {MT}. Gravitational filtering is used to remove the $|F=2,m_F=1\rangle$ atoms to insure the loss rate measurements correspond to a single state and trap depth. %Because atoms in these two states experience different trap depths and retaining a mixture would complicate the interpretation of the loss rate measurements, we use gravitational filtering to remove the atoms in  {the} $|F=2,m_F=1\rangle$ state.  
We set the axial magnetic field to 27 G/cm and hold the atoms for 200 ms so that only those atoms in {the} $|F=2,m_F=2\rangle$ state are retained in the MT.  Atoms in the $|F=2,m_F=1\rangle$ state fall out of the trap as this gradient is insufficient to compensate the gravitational force.  After this quantum state purification step, we then increase the axial magnetic field gradient from 27 G/cm to 272 G/cm in 10 ms and the atoms are held for a period $t$. For trap depths smaller than 1.0 mK, we only increase the axial magnetic field gradient to 81.6 G/cm so that the trap volume explored by the sensor atoms is larger and thus the intra-trap two-body collision rate  {-- which scales with the sensor atom density -- }is minimized.  For this study of the trap depth dependent loss rate of Rb sensor atoms exposed to a H$_2$ background gas, we specifically chose to trap the $|F=2, m_F = 2\rangle$ state to access the largest trap depths possible given our setup.  

As in Refs.~\cite{Booth2019,Shen_2020}, at the end of this hold time, a radio frequency (RF) field, whose frequency is swept from a high frequency to a lower frequency cutoff $\numin$, is turned on to eject all atoms above a certain energy from the trap.  This step sets the trap depth for each loss rate measurement.  The remaining atoms in the MT below this cutoff energy are recaptured in the 3D MOT using the same settings as used for the initial MOT loading, and the signal, $V_{\rm{MT}}$, is recorded. We compute the recaptured fraction, $\fr=V_{\rm{MT}}(t)/V_{\rm{MOT}}$, where $V_{\rm{MOT}}$ is the photodiode voltage just before the MOT to MT transfer.  The quantity $V_{\rm{MOT}}$ is proportional to the atom number transferred into the MT, and this procedure reduces the effects of shot-to-shot variations in the number of atoms loaded into the MOT \cite{Shen_2020}. Following the methods described in  {Ref.}~\cite{Stewart_2022}, we eliminate the contributions from the intra-trap two body collisions and verify that the recaptured fraction versus hold time follows a one-body exponential decay law. Then, we use the method of two-point measurements described in Ref.~\cite{Jones1996} to optimize the determination of the decay constant by placing half of the sample points at the shortest hold time and the remaining half at 1.28 $\tau$, where $\tau$ is the lifetime of the cold sensor atoms. Using this two-point measurement scheme, we obtain a smaller statistical uncertainty in the decay rate than when using a sample set spaced linearly in time. An example data set showing the decay rate measurements using this two point measurement scheme is shown in the inset of  {Fig.}~\ref{fig:RbH2}. In this scheme, we have the following expressions for the recaptured fractions at each hold time, 
\begin{eqnarray}
    \begin{cases}
      f_j(t_j)=f_0\exp(-\Gamma_{\rm{exp}} t_j)\\
%        \nonumber
      f_k(t_k)=f_0\exp(-\Gamma_{\rm{exp}} t_k),
  \end{cases}
  \label{eq:recapftwo}
\end{eqnarray}
where $t_j$ is the first hold time and $t_f$ is the second hold time. The recaptured fraction at zero time is $f_0$, and $f_j$ and $f_k$ are the recaptured fractions at the times $t_j$ and $t_f$, respectively. The experimentally measured loss rate is a sum of the collision-induced loss due to H$_2$ gas ($\gloss$) introduced by heating the NEG, and the constant baseline loss rate, $\Gamma_0$. Namely, $\gexp=\gloss+\Gamma_{0}$. Here $\Gamma_0$ includes all other sources of sensor atom loss including those due to the residual background vapor present before the NEG is heated and to other loss processes not related to background gas collisions \cite{Booth2019,Shen_2020}.

We performed the recapture fraction measurements at the two hold times sequentially in order to minimize the influence of fluctuations in the MOT to MT transfer that affect $f_0$. This procedure allows us to make the assumption that the zero time recapture fraction is the same for each two-point measurement pair. Using this assumption and the log of the ratio of the expressions in Eq.~\ref{eq:recapftwo}, we find,
\begin{eqnarray}
        &\ln\Big[\frac{f_j(t_j)}{f_k(t_k)}\Big]=\gloss (t_k-t_j)+\Gamma_0 (t_k-t_j) \nonumber\\
        \hspace{-10pt}&=\frac{\svlu\Big[(P_k-P_0)t_k-(P_j-P_0)t_j\Big]}{\ig \kb T}+\Gamma_0 (t_k-t_j).
        \label{eq:twopointsolu}
\end{eqnarray}

 {We have} used Eq.~\ref{pressurestd} to rewrite the loss rate due to the added gas on the RHS as $\gloss^{(j)} = \Ng^{(j)} \svlu = \frac{(P_j-P_0)}{\ig \kb T} \svlu $.  Here $P_j$ and $P_k$ are the pressure readings from the (uncalibrated) IG for the measurements corresponding to the holding times $t_j$ and $t_k$, respectively.  The IG response to the added H$_2$ gas is obtained by subtracting the IG pressure readings at the background level, $P_0$, from the measured pressures $P_j$ and $P_k$.  (Note that these pressure readings need not be accurate. The only requirement of the IG is that its response is linear in the gas pressure.) The H$_2$ calibration factor of the ion gauge is $\ig$, and $T$ is the ambient temperature recorded by a thermal couple gauge, which is the same for different hold times.  We repeat this two point measurement procedure for many different trap depths and thus map out the function of $\svlu/\ig$ versus the trap depth.

The last term in Eq.~\ref{eq:twopointsolu} quantifies the contributions from loss mechanisms that do not depend on the density of the H$_2$ gas species introduced and can be obtained from the pre-heating (baseline) loss rate measurements.  That is, the pressure of the ambient gas $P_0$ and the baseline loss rate for a variety of trap depths $\Gamma_0(U)$ are determined before any H$_2$ gas is introduced into the system.  For this work, $\gloss$ is two orders of magnitude higher than $\Gamma_0$.

\subsection{Device  {(B)}}

This apparatus was used to perform loss rate measurements of both Rb and Li atoms exposed to the same H$_2$ background gas.  The aim was to determine the ratio of the total cross sections, 
\begin{equation}
     {R = \frac{\svtli}{\svtrb}.}
     \label{eq:Rlirb}
\end{equation}  
For these measurements, the Li and Rb atoms were confined in very shallow traps to ensure that the measured loss rate is approximately equal to the total collision rate (i.e.~$\gloss \approx \gtot$).  In this limit, the cross section ratio can be obtained by inspecting the ratio of the Li and Rb loss rates.  In addition, the exact value of the H$_2$ gas density need not be known.  All that is required is that the H$_2$ gas density be the same for the Rb and Li loss rate measurements.  To extract the cross section for collisions with only H$_2$ molecules, loss rate measurements of Rb and Li were made in pairs where the H$_2$ gas density was the same for each pair.  Then, as described below, these pair measurements were made at a variety of different H$_2$ gas densities.

In  {device} (B),  {effusive ovens emit thermal, parallel atomic beams of the two species -- Rb and Li -- which are subsequently decelerated and cooled in a Zeeman slower region by a counter-propagating bi-chromatic optical beam.}
%parallel, thermal, atomic-beams of the two species, Rb and Li, emerging from a pair of effusive ovens are decelerated and cooled in a Zeeman slower region by a counter propagating bi-chromatic optical beam. 
The atoms are then captured in a 3D MOT and transferred into a co-located  {MT}.  The 3D MOT region vacuum is maintained by an ion pump and a {NEG}  and is separated from the oven region by a differential pumping tube.  The NEG is the same model as the one used in {device} (A). In order to increase the H$_2$ pressure above its baseline pressure while keeping the other gasses relatively constant, we applied a voltage across the NEG within the same range as in {device} (A). Since we are interested in the loss rate ratio of the Li and Rb sensor ensembles when exposed to the same H$_2$ gas, we could have co-trapped both sensor ensembles in the same {MT} and simultaneously measured the Rb and Li loss rates.  However, this was not done because the optimal operating parameters for slowing, trapping and transferring atoms from the MOT to the MT are significantly different for Rb and Li.  Instead, we interlaced the loss rate measurements for the two species as described below, in order to to minimize systematic errors associated with pressure changes during the measurements.  As in device {A}, we measured the atom numbers by observing the atomic fluorescence signal of the atoms in the MOT using a photo-detector.  We verified the linearity of our photo-detector signal with atom number by direct comparison with the atom number determined by absorption images of our atoms (using a different detector) over a large range of atom numbers.  The atomic fluorescence signal was used for this study because it has a much higher sensitivity than the absorption image method.

For the Lithium measurements, the MOT pump laser was detuned by $\dpump=50$~MHz below the D$_2$ (2$^2$S$_{1/2}$ $\rightarrow$ 2$^2$P$_{3/2}$) $F=3/2\rightarrow F'=5/2$ transition of the $^{6}$Li sensor atoms.  After a MOT loading sequence of 10~s, % {\textbf{[Could we find a different notation other than [ ] in this sentence? It looks like a citation $\rightarrow$]}}$
$(1-2)\times10^7$  cooled Li atoms reside in the MOT, corresponding to a fluorescence signal of $\vmotli  \sim (1.4-2.8)$~V.  After the MOT is loaded, we extinguish the Zeeman slowing optical beam and block the thermal atomic beams using a mechanical beam block.  We then initiate a 100~ms ramp of the quadrupole magnetic field gradient of the MOT from 10~G/cm to 30~G/cm, followed by a cooling sequence for which we change the detuning of the pump and repump laser to $\dpump = -13.6$~MHz and $\drepump = -9$~MHz. After this 4~ms of cooling, the Li atoms are optically pumped to the $F=1/2$ state in 10~$\mu$s by extinguishing the repump light before the pump light.  Atoms in the $|F=1/2, m_F = -1/2\rangle$  {state} are low field seeking and are trapped in the MT given their total energy  {is less than the maximum potential energy, $\umax/\kb = 300 \; \mu$K.}  This maximum potential energy is a fundamental limit  {resulting from the} magnetic moment sign change of the $|F=1/2, m_F = -1/2\rangle$ state at a field of approximately 30~G \cite{Shen_2020}.  Because of the temperature of atoms in the MOT (on the order of 1mK) and the limited trap depth, we typically transfer only around 2\% ([2-4]$\times10^5$) of atoms from the MOT to the MT.  After transfer, we hold the atoms for a time $t$.  After this hold time, we recapture those atoms remaining in the MT into the 3D MOT using the same settings as used for the initial MOT loading.  As described for  {device} (A), we record the MOT fluorescence after recapture and compute the recaptured fraction, $\fr=V_{\rm{MT}}(t)/V_{\rm{MOT}}$, where $V_{\rm{MOT}}$ is the photodiode voltage just before the MOT to MT transfer.  No radio-frequency radiation is applied to set the trap depth for the  {Li} measurement.  Rather, we rely on the fundamental trap depth limit described above for the $|F=1/2, m_F = -1/2\rangle$ state.

For the {Rb} measurements, the 3D MOT light operates on the same transition as  {device} (A), but the experimentally optimal values differ slightly.  During loading, the trapping laser is detuned to $\dpump=-20$~MHz, below the pump transition.  After loading the MOT for 2~s, we observed a fluorescence signal of $\vmotrb = (1.6-3.2)$~V, corresponding to  $(0.8-1.6)\times10^7$ Rb atoms.  After the MOT was loaded, we further cooled the trapped ensemble for 15~ms by detuning the pump light to $\dpump=35$~MHz below the $F=2\rightarrow F'=3$ transition.  The repump light is extinguished for the last 10~ms of this procedure to optically pump the atoms into the $F=1$ hyperfine manifold.  We then extinguished all light and increased the quadruple magnetic field axial gradient from 20~G/cm to 60~G/cm and held the atoms in the $|F=1, m_F = -1\rangle$ state for a time $t$.  This MOT to MT transfer procedure is similar to that for {device} (A).  The principal difference is that the atoms in  {device} (B) are trapped in the $|F=1, m_F = -1\rangle$ state and thus no spin polarization beam or gravitational filtering was required since this is the only trapped state for the $F=1$ manifold.  Immediately after the atoms are transferred into the MT and then immediately before recapture of the atoms into the MOT (after the hold time of $t$ seconds), we apply a radio frequency (RF) field for a duration of 100~ms and whose frequency is swept from a high value to a lower value to eject all atoms above a certain energy from the trap.  We chose the RF field parameters to set the magnetic trap depth to $U \le 200~\mathrm{\mu K}$.  As with  {device} (A), atoms remaining in the MT below this cutoff energy are recaptured in the 3D MOT.  The fluorescence is then recorded, and the recaptured fraction is found.

\section{Discussion of Results}

\subsection{Trap loss measurements of Rb exposed to H$_2$}
\label{sec:univ-RbH2}

Using apparatus A, we performed an extensive study of the trap loss rate versus trap depth of $^{87}$Rb atoms exposed to a H$_2$ background gas over a much wider range of depths than previously studied \cite{Shen_2020}.  In particular, we made loss rate measurements from 0.5~mK to 6~mK, in an effort to define the shape of the loss rate versus trap depth curve more precisely and facilitate a clearer comparison with the universal prediction. However, as the trap depth increases, the collision-induced heating of the trapped ensemble becomes more pronounced, confounding the loss rate measurements. %since the fraction of collisions that do not eject sensor atoms increases.  
Here we performed additional measurements and data analysis to properly account for this effect.  

Collisions that fail to eject sensor atoms from the trap change the sensor ensemble energy distribution over time \cite{Beijerinck2000,Beijerinck2000Nov}.  Because the effective trap depth depends on the sensor-atom energy distribution in the trap, this results in a time-dependent and monotonically decreasing effective trap depth.  If not accounted for, this leads to a systematic overestimate of the measured loss rate with trap depth and, consequently, a systematic underestimate in the deduced $\svt$ value.

\begin{center}
\begin{figure}[ht!]
\includegraphics[width=0.5\textwidth]{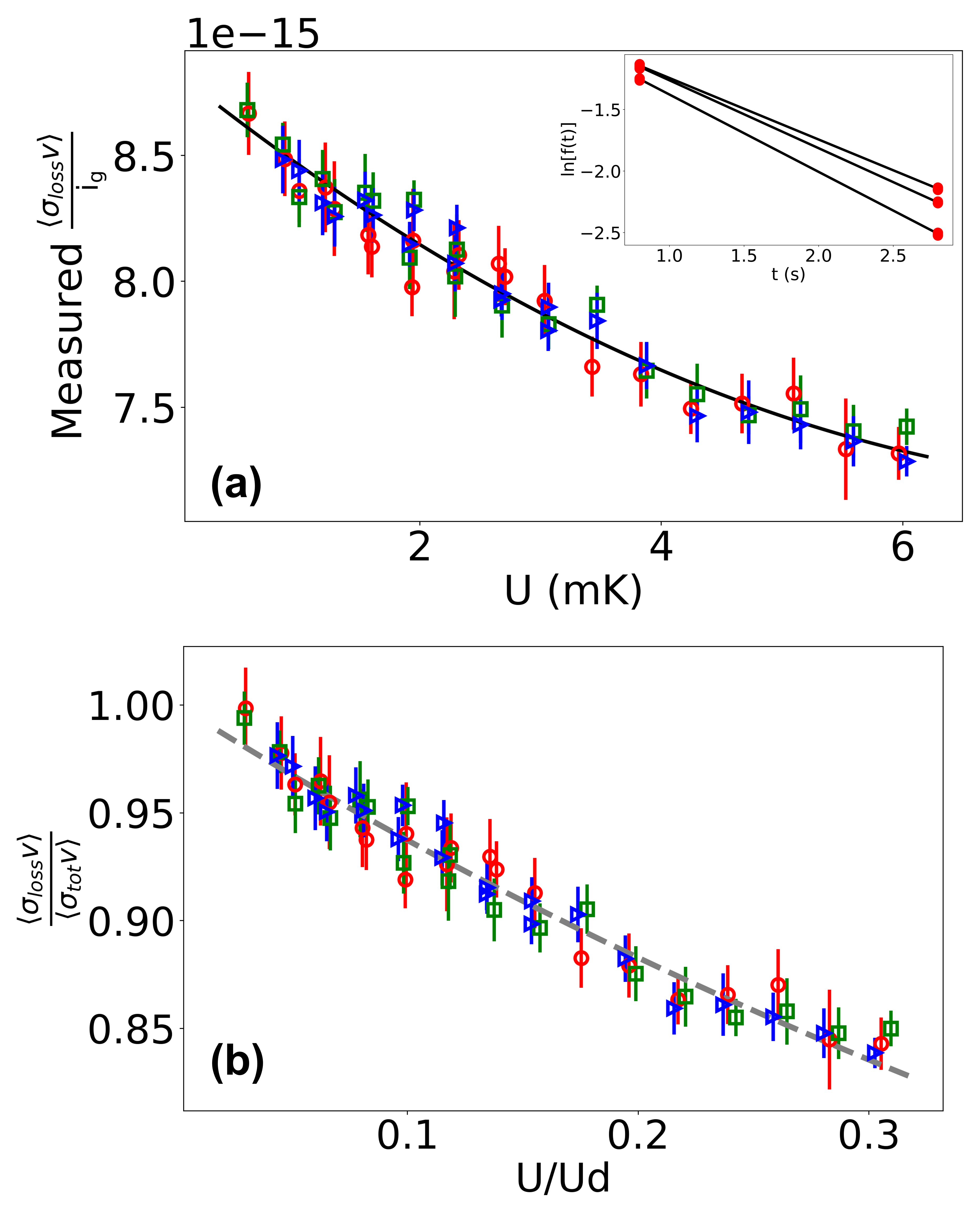}
\caption{ Measured trap loss rate coefficients as a function of trap depth for $^{87}$Rb atoms held in a magnetic trap and exposed to a H$_2$ gas.  The data in (a) shows the value of $\svlu/\ig$ as a function of the trap depth after accounting for sensor ensemble heating (see appendix). Different colors represent data taken on three different calendar dates. Each datapoint is the average of three two-point decay rate measurements. Examples of two-point decay measurements, each at a different trap depth, are shown in the inset. The black curve in (a) is the second order polynomial fit to the data provided in Eq.~\ref{eq:svlufit} and used to extrapolate the value at $U=0$.  By instead fitting the same data to the universal function, Eq.~\ref{eq:svl_universal}, we determine the value of $\svt$ and $\ig$ for the data taken on three different dates.  Fitting the red circles we obtain $\svt=5.49(6)  \times 10^{-15} \; \mathrm{m}^3/\mathrm{s}$ and $\ig=0.632(7)$ . For the blue triangles, we have $\svt=5.54(7)  \times 10^{-15} \; \mathrm{m}^3/\mathrm{s}$ and $\ig=0.625(7)$. For the green squares, we have $\svt=5.58(6)  \times 10^{-15} \; \mathrm{m}^3/\mathrm{s}$ and  $\ig=0.645(10)$.  The universal fit to the combined set of data yields $\svt=5.52(9) \times 10^{-15} \; \mathrm{m}^3/\mathrm{s}$ and $\ig=0.633(10)$.  The universal plot of all the data (the loss rate coefficient is normalized by the value at zero trap depth and the trap depth is re-scaled by $\ud$) is shown in (b), and the grey dashed line represents the universal fit function.}
\label{fig:RbH2}
\end{figure}
\end{center}

Although the fitting uncertainty for $\svt$ decreases as the range of trap depths increases, the systematic error caused by unaccounted sensor-ensemble heating will increase with trap depth. We have found that the ensemble heating error becomes dominant once the scaled trap depth, $\uud$, exceeds 0.3 \cite{Shen_2020}. Therefore, the uncertainty in determining $\svt$ is limited by ensemble heating. To resolve this problem and eliminate the resulting systematic uncertainty, we apply a correction to the trap depth by measuring the energy distribution of the ensemble as a function of the hold time.  This allows us to infer the correct effective trap depth as a function of the hold time (details are provided in Appendix~\ref{app:heatingcor}). 
After properly accounting for ensemble heating, we obtained corrected values of $\svlu/\ig$ as a function of the trap depth. For each trap depth, we repeated the measurements for three times and we repeated the whole measurement sequence on three different calendar dates.  The data are shown in Fig.~\ref{fig:RbH2}(a). The overlap of the data on the three different dates indicates the reproducibility of the measurements. 

One goal of this measurement is to determine the precise shape of $\svlu$ as a function of the trap depth without any assumption that it follows a particular curve.  For this purpose, we fit the combined data on different dates to a second order polynomial.  This fit is also used to extrapolate and find the value of $\svlu/\ig$ at zero trap depth,  {which is} needed to normalize the loss rate data as shown in Fig.~\ref{fig:RbH2}(b).  The best fit to the data is
\begin{eqnarray}
        \frac{\svlu/\ig}{1\times 10^{-15}\rm{m}^3/\rm{s}}&=&8.82(6)-0.38(4)\left(\frac{U}{\rm{mK}}\right) \nonumber \\
& &\hspace{28pt}+\  0.022(6)\left(\frac{U}{\rm{mK}}\right)^2.
\label{eq:svlufit}
\end{eqnarray}
Another goal of this measurement is to determine the value of $\svt$ and $\ig$ simultaneously by fitting these data to the universal function, Eq.~\ref{eq:svl_universal}. As expected, the fitted values on different dates agree with each other.  We find $\svt=[5.49(6), 5.54(7),5.55(8)]\times 10^{-15}$~m$^3$/s, and
$\ig=[0.632(7), 0.625(7),0.645(10)]$. The universal function fit to the combined data set is shown in Fig.~\ref{fig:RbH2}(b).  The (heating-corrected) values obtained by fitting the combined data set are 
$\svt=5.52(9)\times \ 10^{-15}$ m$^3$/s and $\ig=0.633(10)$.

% see here for "Analysis of the Iterative Heating Fraction Fitting Procedure" https://qdg-forum.phas.ubc.ca/t/pat-experiment-daily-logs-2021/354/74

% https://qdg-forum.phas.ubc.ca/t/pat-experiment-daily-logs-2021/354/73

\subsection{Cross species sensor atom comparison}

In order to determine the Rb+H$_2$ cross section independently from that determined by the universality fit, we used device  {(B)} to perform loss rate measurements of trapped Rb and of trapped Li atoms exposed to the same H$_2$ background gas pressure.  By confining the atoms in very shallow magnetic traps where $\gloss \approx \gtot$, we were able to determine the ratio of the Li+H$_2$ and Rb+H$_2$ cross sections, $\svtli:\svtrb$.  Using this measured ratio and the \textit{ab initio} computed value for the Li+H$_2$ cross section, we were able to determine the Rb+H$_2$ cross section and compare this to that found using the universality fit described in Sec.~\ref{sec:univ-RbH2}

\begin{center}
\begin{figure}[ht!]
    \includegraphics[width=\columnwidth]{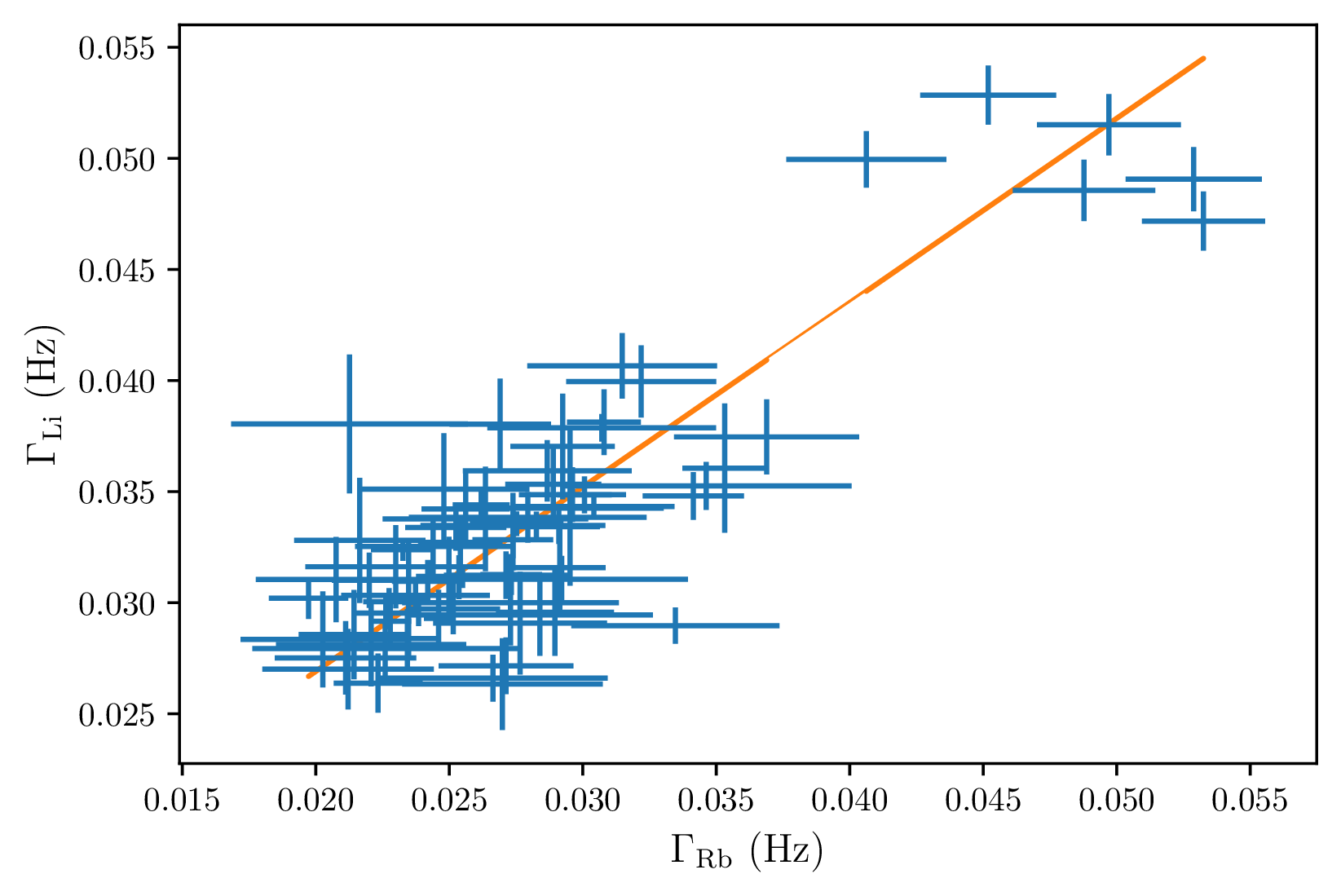}
\caption{Trap loss rate measurements for an ensemble of $^6$Li and $^{87}$Rb atoms exposed to the same H$_2$ background gas produced by heating the NEG.  As the H$_2$ pressure increased, the loss rates increased for both ensembles.  Here the loss rates are plotted as the ordered pair $(\grb,\gli)$, and the data is fit to a line whose slope provides the ratio of the loss rates and thus the ratio of the cross sections, $R = \frac{\gli}{\grb} = \frac{\svtli}{\svtrb}$.  For these data, we find $R = 0.83 \pm 0.05$.
}
\label{fig:RbLiH2}
\end{figure}
\end{center}

For these measurements, the aim was simply to determine the ratio of the total cross sections.  Therefore, the exact value of the H$_2$ gas density need not be known.  In fact, all that is required is that the \emph{change} of the H$_2$ gas density be the same for each pair of measurements of the Rb and Li trap loss rates, $\grb$ and $\gli$.
For traps of zero depth with respect to collisions with H$_2$ molecules ($\gloss \approx \gtot$), we have
\bea
\grb & = & \Ng \svtrb + \grbz\\
\gli & = & \Ng \svtli + \gliz
\eea
where $\grbz$ and $\gliz$ are the Rb and Li baseline loss rates measured before the NEG is heated. As for  {device (A)}, $\Ng$ is the additional number density of H$_2$ gas resulting from heating the NEG.  As the NEG was heated, $\Ng$ increased, and measurements of $\grb$ and $\gli$ are made in pairs with minimum time between the two measurements so that the value of $\Ng$ is the same for each data pair.  In Fig.~\ref{fig:RbLiH2} we plot the ordered pairs $(\grb,\gli)$ for different values of $\Ng$ in a 2D scatter plot where the abscissa is the Rb loss rate and the ordinate is the Li loss rate. We then perform an orthogonal distance regression (ODR) fit of the data to a line.  
The slope of this line is the ratio:
\bea
R = \frac{\svtli}{\svtrb}.
\eea

From the data, we found $R = 0.83 \pm 0.05$.  Based on the theoretically predicted value from Ref.~\cite{PhysRevA.99.042704,PhysRevA.105.039903} of $\svtli= 3.13(6) \times 10^{-15}$~m$^3$/s, we can infer that $\svtrb = 3.8(2) \times 10^{-15}$~m$^3$/s.  
%Also, we need to address the change due to the trap depth.   The maximum depth was X for each sensor atom set by this and that.  For Li, this was the zero trap depth limit and for Rb, the maximum suppression at this depth was less than 4\%. 
% Erik: I address this below

% computed values : Li = 3.1011 / Rb = 3.5067 
% should we account for trap depth

Due to its low mass and the small maximum trap depth ($\umax/\kb = 300 \; \mu$K) for the $|F=1/2, m_F = -1/2\rangle$ state, Li is always in the zero trap depth limit with $\frac{U}{U_d}\simeq0.1$\%.
For Rb, we use radio-frequency radiation (as describe above) to limit the magnetic trap depth to $U<200~\mathrm{\mu K}$, for which the deviation of $\svloss$ from $\svtot$ is less than 0.1\% as per Eq.~\ref{eq:svlufit}.  Thus we expect trap depth effects to be negligible compared to our statistical uncertainty.

% Erik: I used the Shen_2020 paper to get the beta coefficients, and used U/Ud = 200uK/21.5mK 
% to find
% sigma_lossv = 0.993 sigma_totv

\subsection{\textit{Ab Initio} calculation predictions}

In addition to  {experimentally} investigating the  {rate coefficient ratio} $R = {\svtli}:{\svtrb}$, we have also computed it.  We find an \emph{ab initio} prediction of $R \approx   {0.869}$, which is within the range of experimentally measured value.  Specifically, we have performed coupled channel (CC) calculations to find the Rb+H$_2$ and Li+H$_2$ cross sections.  The details of these calculations is provided in the appendix.

In brief, the Rb+H$_2$ cross section was calculated using a PES  kindly provided to us from the authors of Ref.~\cite{Upadhyay2019}.  This PES was obtained by \emph{ab initio} calculations with the unrestricted coupled cluster method with single, double and perturbative triple excitations [UCCSD(T)] \cite{Deegan1994}.  The Li+H$_2$ cross section was found using a PES published in Refs.~\cite{PhysRevA.99.042704,PhysRevA.105.039903}.  To validate the accuracy of our coupled channel calculations, we first recomputed the rate coefficient for Li + H$_2$ collisions.  We found a value of $\svtli = 3.101(6)\times 10^{-15}$~m$^3$/s, in good agreement with the value of $\svtli = 3.13(6)\times 10^{-15}$~m$^3$/s from Refs.~\cite{PhysRevA.99.042704,PhysRevA.105.039903}. We then computed the cross sections and rate coefficients for Rb + H$_2$ collisions using the PES from \cite{Upadhyay2019}, yielding $\svtrb = 3.60(7)\times 10^{-15}$~m$^3$/s.  Using these \textit{ab initio} values, we predict a cross section ratio of $R = 0.869$ in good agreement with the experimental value of $R=0.83(5)$.

Although the accuracy of the Rb+H$_2$ PES has not been independently validated, the computed cross section based on this PES is in good agreement with the value found experimentally by comparison with Li.  Since both the experimental value and theoretical estimate for the Rb+H$_2$ cross section agree with each other within their respective error bars and are inconsistent with the value extracted from fitting the Rb loss rate curve given the universal function (approximately 50\% larger), this indicates that the cross sections for Rb+H$_2$ and Rb+He collisions previously determined using the universality result are systematically overestimated \cite{Booth2019,Shen_2020,Shen_2021}.

\section{Cross-Species Calibration}
The application of cold atom trap losses to pressure measurements is an exciting realization of a fully quantum pressure sensor. The key ingredient for this sensor is the determination of the velocity-averaged total collision cross-section between the trapped species and the background gas, $\svt$. There are three methods for determining these quantities: calibration of the loss rate against a known pressure, \textit{ab initio} quantum scattering computation of $\svt$, or relying on the universality description of the quantum diffractive collisions. No single method can span the entire range of collision partners. Calibration against a known background pressure can be implemented for a limited number of inert gas species where the pressure can be determined independently using an orifice flow standard. The \textit{ab initio} QS computation technique may be limited to collisions between light species by the computation complexity associated with accurately finding the PES for many-electron systems. The universality approach, as shown here, is inaccurate for collisions between low reduced mass and low polarizability collision partners. To overcome its limitations, as we have demonstrated in this study, we leveraged a cross-species calibration for the loss rate coefficients. While we have calibrated $\svtrb$ to the \textit{ab initio} value of $\svtli$, the inverse method can readily be applied. Namely, when low mass species such as Li are used as sensor atoms, it is unlikely that the quantum diffractive universality description will be applicable to any collision induced loss measurements, regardless of the collision partner. The strength of using trapped Li sensor atoms, the ability to perform precise QS computations of $\svt$, is realized for only a handful of particles. However, the Li cross sections can be calibrated against those for a heavy sensor atom such as Rb, where the QDU technique is available for a wide range of background gas collisions, including other significant vacuum constituents such as N$_2$ and Ar. In short, cross-species calibration allows us to exploit the strengths of each sensor atom to effect an optimum quantum pressure sensor/standard.

\section{Conclusions}\label{sum}

The main goal of this work has been to demonstrate, experimentally, an anticipated deviation of the total collision cross section for light collision partners from the predictions of the universal $\PQDU$ function \cite{Booth2019, Shen_2020, Shen_2021}. We have carefully characterized the loss rate coefficient, $\svloss$, as a function of trap depth for Rb-H$_2$ collisions and fit these data to the universal prediction. This yielded a value of $\svtrb = 5.52(9)\times 10^{-15}$~m$^3$/s.  Next, we implemented a ratiometric technique which compared the loss rates of Li atoms and Rb atoms, confined in very shallow magnetic traps, simultaneously exposed to background H$_2$. The choice of Li for cross-species calibration is motivated by the simplicity of Li - H$_2$ interactions, which makes highly accurate predictions of Li + H$_2$ scattering cross sections feasible. The \textit{ab initio} predictions of cross sections for Li + H$_2$ collisions at room temperature were previously published by other authors \cite{PhysRevA.99.042704,PhysRevA.105.039903}. To ensure the accuracy of our results, in the present work, we have recomputed the rate coefficients for Li + H$_2$ collisions, $\svtli = 3.101(6)\times 10^{-15}$~m$^3$/s, in good agreement with the value of $\svtli = 3.13(6)\times 10^{-15}$~m$^3$/s from Refs.~\cite{PhysRevA.99.042704,PhysRevA.105.039903}. Under our experimental conditions, the loss rate ratio is $R = \svtli : \svtrb = 0.83(5)$, yielding $\svtrb = 3.8(2)\times 10^{-15}$~m$^3$/s when calibrated to the Li-H$_2$ total collision cross section value. We have also computed the cross sections and rate coefficients for Rb + H$_2$ collisions using the PES from \cite{Upadhyay2019}, yielding $\svtrb = 3.60(7)\times 10^{-15}$~m$^3$/s. Thus, the \textit{ab initio} value of $R = 0.869$ agrees with the experimentally determined value and verifies that the experimental value for $\svtrb$ deviates from the value derived from the universal function, which would predict $R = 0.567(3)$.

The universality of quantum diffractive collisions was previously exploited to demonstrate self-calibrating, primary vacuum pressure standard based on trapped Rb atoms \cite{Booth2019,madison2018,Shen_2020,Shen_2021}. 
The energy distribution function $\PQDU$ was shown to be universal for collisions of neutral species because diffractive collisions are predominantly determined by the long-range part of PES and because the dependence of the total scattering cross section on short-range interactions is removed by Maxwell-Boltzmann averaging. However, both the relative contribution of long-range glancing collisions and the effect of Maxwell-Boltzmann averaging are expected to diminish for species with small mass and, consequently, small polarizability. 

The present work demonstrates this expected departure of Rb + H$_2$ diffractive collisions from universality. We have shown that this limitation can be overcome by performing a cross-calibration of $\svtrb$ to $\svtli$. However, we recognize that there is a limited number of collision partners which are amenable to this \textit{ab initio} approach. This has several important implications: 
First, we believe that heavy polarizable atoms may be a preferred choice of pressure sensors. These can be calibrated for a wide range of collision partners via the universality method while light collision partners, such as H$_2$ and He, can readily be calibrated against other sensor species, such as Li, for which $\svt$ can be computed accurately. Second, the inverse method, namely a lower mass, lower polarizability sensor atom, like Li, can be calibrated to a heavier mass sensor atom (such as Rb) for a wide range of collision partners accessible to the universal quantum diffractive collision formalism.
%Second, our work shows that even pressure sensors based on heavy trapped atoms may require calibration for measurements of pressure of He and H$_2$ gases. 
%For such species, it may be necessary to obtain information on the collision cross sections between the trapped species and the atoms/molecules in the ambient gas from independent sources. This can be done by cross-species calibration, as demonstrated in this work, or by precise quantum scattering calculations. 
%\textcolor{blue}{Third, we have demonstrated how to calibrate the loss rate coefficient, $\svt$, from one sensor species against another. This greatly expands the functionality of pressure sensors based on cold atom trap loss.}
Finally, our work brings to light the necessity of assessing the limitations of universality of quantum diffractive collisions for light background species as well as light sensor atoms. 

\section{Contributions}

Author contributions are as follows.  P.S., A.D., and R.S. took and analyzed the high trap depth loss rate measurements for Rb expoed to H$_2$.  E.F. and D.U. took and analyzed the combined Rb+H$_2$ and Li+H$_2$ measurements.  R.V.K. and  {K.R.H.} performed and analyzed the Rb+H$_2$ and Li+H$_2$  {coupled-channel quantum scattering} calculations.  P.S. and J.L.B. performed and analyzed preliminary, single-channel, quantum scattering calculations.  K.W.M. and J.L.B. obtained funding for the research and conceived and planned the study.  K.W.M. wrote the manuscript with contributions from the other co-authors.

\begin{acknowledgments}

We acknowledge financial support from the Natural Sciences and Engineering Research Council of Canada (NSERC/CRSNG) and the Canadian Foundation for Innovation (CFI). This work was done at the Center for Research on Ultra-Cold Systems (CRUCS). P.S. acknowledges support from the DFG within the GRK 2079/1 program.  {K.R.H. acknowledges financial support from the Quantum Electronic Science \& Technology (QuEST) Award given by the Stewart Blusson Quantum Matter Institute.}

\end{acknowledgments}

\appendix

\section{Heating rate corrections}\label{app:heatingcor}

\begin{center}
\begin{figure}[ht!]
\includegraphics[width=0.45\textwidth]{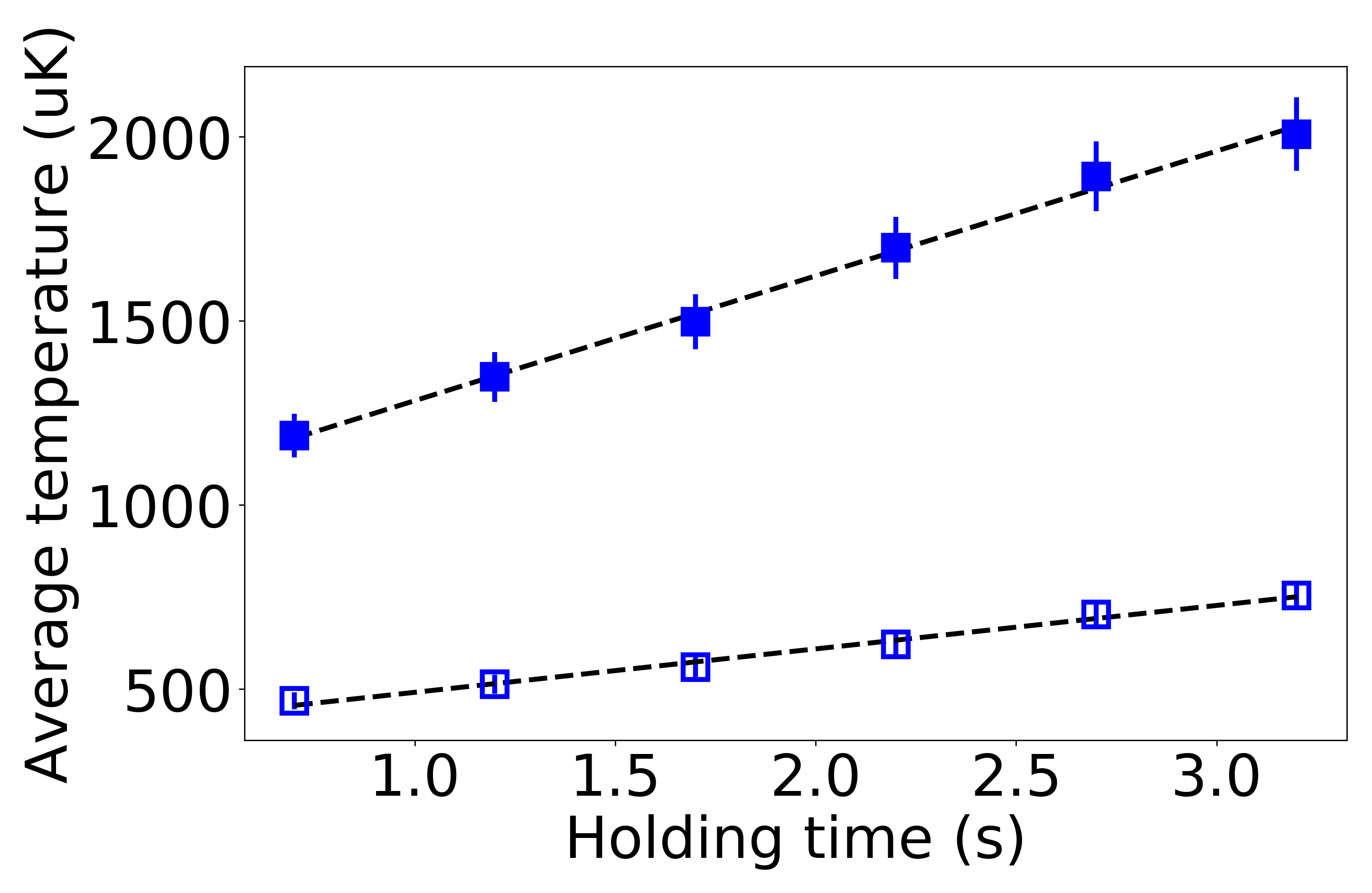}
\caption{The measured temperature of the sensor ensemble versus the holding time for atoms trapped in $F=2$ state at the magnetic trapping current, $I=60$ A (open blue squares) and  $I=200$ A (blue filled squares). The ensemble temperature is observed to increase linearly with the holding time. The dashed lines are linear fits to the data, $T(t)=T_0+mt$, with $T_0 = 377(26), 944(69) \; \mu$K and $m= 115(14), 339(36) \; \mu$K/s respectively.}
\label{fig:tempvstime}
\end{figure}
\end{center}

In this section, we describe how ensemble heating is accounted for in our determination of $\svt$.
For clarity, we repeat here the expression in Eq.~\ref{eq:twopointsolu} describing the fraction of 
sensor atoms remaining in the MT at times $t_j$ and $t_k$.
\begin{eqnarray}
        &\ln\Big[\frac{f_j(t_j)}{f_k(t_k)}\Big]=\gloss (t_k-t_j)+\Gamma_0 (t_k-t_j) \nonumber\\
        \hspace{-30pt}&=\frac{\overline{\svlu}\Big[(P_k-P_0)t_k-(P_j-P_0)t_j\Big]}{\ig \kb T}+\Gamma_0 (t_k-t_j).
\end{eqnarray}
Here, we have replaced $\svlu$ with $\overline{\svlu}$, indicating that the rate coefficient, modeled by the universal function in Eq.~\ref{eq:svl_universal}, is averaged over the energy distribution of the ensemble, $\rho(E)$ \cite{Shen_2021}.  This accounts for the trap depth of each atom in the ensemble with total energy $E \ne 0$ being different than the maximum trap depth, $E_{\mathrm{max}}$.
\begin{eqnarray}
\overline{\svlu} = \frac{\int_0^{E_{\rm{max}}}\svlE\rho(E) dE}{\int_0^{E_{\rm{max}}}\rho(E) dE}.
\end{eqnarray}
We can now include collision-induced ensemble heating by explicitly writing the ensemble energy distribution as a time dependent function, $\rho(E,t)$, and evaluating the integrated loss rate coefficient from zero time to the measurement time $t$.  In this case, the ratio of the remaining atoms satisfies
\begin{eqnarray}
    \ln\Bigg[\frac{f_j(t_j)}{f_k(t_k)}\Bigg]-\Gamma_0 (t_k-t_j)\nonumber \\
    & &\hspace{-120pt}=\frac{(P_k-P_0)}{\ig \kb T}\int_0^{t_k}\frac{\int_0^{E_{\rm{max}}}\svlEt\rho(E,t) dE}{\int_0^{E_{\rm{max}}}\rho(E,t) dE}dt \nonumber\\
    & &\hspace{-115pt}-\frac{(P_j-P_0)}{\ig \kb T}\int_0^{t_j}\frac{\int_0^{E_{\rm{max}}}\svlEt\rho(E,t) dE}{\int_0^{E_{\rm{max}}}\rho(E,t) dE}dt. \nonumber \\
    \label{eq:twopointtimeintermed}
\end{eqnarray}
To simplify the expression, we can factor out the loss rate coefficient at zero time, $\overline{\svlu}_{t=0}$, which gives the following final expression,
\begin{eqnarray}
    \ln\Bigg[\frac{f_j(t_j)}{f_k(t_k)}\Bigg]-\Gamma_0 (t_k-t_j)\nonumber \\
    & &\hspace{-110pt}=\frac{\overline{\svlu}_{t=0}}{\ig \kb T}\Big[(P_k-P_0) H(t_k)-(P_j-P_0)H(t_j)\Big], \nonumber \\
    \label{eq:finaltwopointsolu}
%    &=&\frac{H_{0,t_j}}{ig}\Bigg[\frac{(P_k-P_0)}{\kb T}\frac{H_{0,t_k}}{H_{0,t_j}}-\frac{(P_j-P_0)}{\kb T}\Bigg],\nonumber\\
\end{eqnarray}
where we define the heating factor as,
\begin{eqnarray}
    H(t')&\equiv&\frac{\int_0^{t'}\frac{\int_0^{E_{\rm{max}}}\svlEt\rho(E,t)dE}{\int_0^{E_{\rm{max}}}\rho(E,t)dE}dt}{\frac{\int_0^{E_{\rm{max}}}\svlE\rho(E,t=0)dE}{\int_0^{E_{\rm{max}}}\rho(E,t=0)dE}}\nonumber \\
    &\equiv&\frac{\int_0^{t'}\frac{\int_0^{E_{\rm{max}}}\svlEt\rho(E,t)dE}{\int_0^{E_{\rm{max}}}\rho(E,t)dE}dt}{\overline{\svlu}_{t=0}}.
    \label{eq:Hfactor}
\end{eqnarray}
In order to evaluate these expressions, we measured the sensor ensemble energy distribution as a function of the hold time.  Specifically, we determined $\rho(E,t)$ from the measured cumulative energy distribution of the atoms in the trap.  This was done by measuring the number of atoms remaining in the trap after ejecting atoms above an energy cutoff using a radio-frequency field as described in Ref.~\cite{Shen_2020}.  We found that the distribution is well approximated by a Maxwell Boltzmann distribution with an energy offset,
\begin{eqnarray}
        \rho(E,t) &=& \nonumber\\
 & &\hspace{-40pt}\Theta(E-E_{\rm{min}}) \cdot 2 \left(\frac{E-E_{\rm{min}}}{\pi}\right)^{\frac{1}{2}} \left(\frac{1}{k_B T(t)} \right)^{\frac{3}{2}} e^{-\frac{E-E_{\rm{min}}}{k_B T(t)}}. \nonumber\\
\label{eq:rhoE}
\end{eqnarray}
We measured the evolution of $\rho(E,t)$, specifically the variation of $T$ and $E_{\mathrm{min}}$ with hold time, at the two different trapping currents used in this study, $I=60$ A and $I=200$ A.  We observed that $E_{\mathrm{min}}$ was constant and that the temperature varied linearly in time as shown in figure~\ref{fig:tempvstime}.  The ensemble temperature is modeled by $T(t)=T_0+mt$.

The experimentally determined ensemble energy distribution can then be used to evaluate the integrals in Eq.~\ref{eq:Hfactor} to determine the heating factor.
Using this, we can solve for the value of $\overline{\svlu}_{t=0}/\ig$, by plugging the computed heating factors into Eq.~\ref{eq:finaltwopointsolu},
\begin{eqnarray}
    \frac{\overline{\svlu}_{t=0}}{\ig \kb T}=
    \frac{\ln\Big[\frac{f_j(t_j)}{f_k(t_k)}\Big]-\Gamma_0 (t_k-t_j)}{\Big[(P_k-P_0)H(t_k)-(P_j-P_0)H(t_j)\Big]}.\nonumber \\
    \label{eq:iteratemethod}
\end{eqnarray}

\begin{center}
\begin{figure}[ht!]
\includegraphics[width=0.45\textwidth]{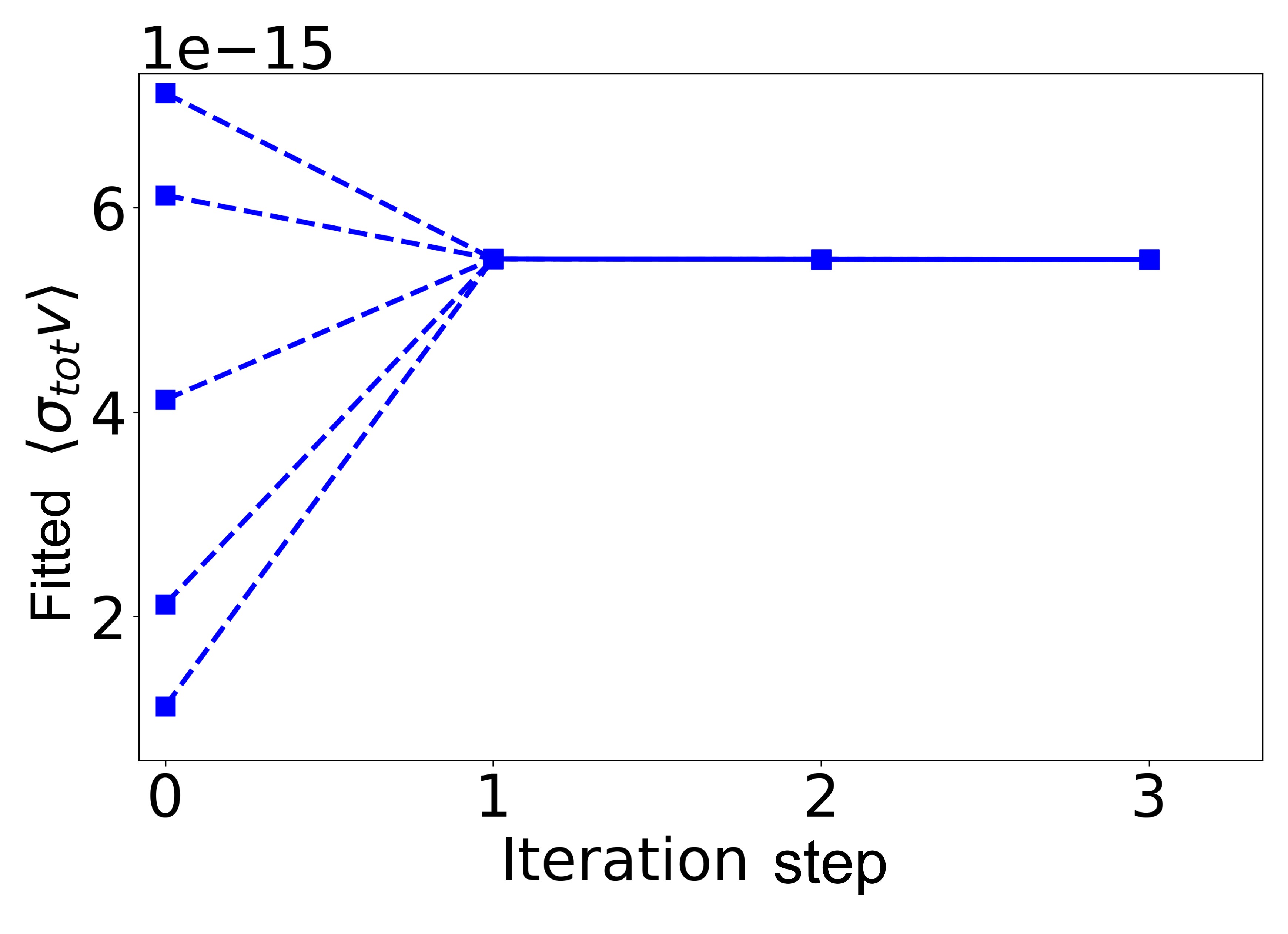}
\caption{Fitted $\svt$ results as a function of the iteration step for atoms trapped in $F=2$ state. The $\svt$ value at step zero represents the value of the initial guess. Here the initial guesses are varied by more than a factor of 6 and yet they all converge to the same $\svt$ value, $5.49\times10^{-15} \rm{m}^3/\rm{s}$.}
\label{fig:iterationmethod}
\end{figure}
\end{center}

However, the right-hand side of Eq.~\ref{eq:iteratemethod} also depends on the unknown parameter, $\svlu$ (included in the heating factor).  This necessitates a self-consistent method to find the solution of Eq.~\ref{eq:iteratemethod}. To achieve this, we used the following iteration fitting method to determine the final $\svt$ and $\ig$ values:
\begin{enumerate}
    \item We pick an initial guess value for $\svt$ to determine a guess for the $\svlu$ function and feed this
     into the expression for $H$, Eq. \ref{eq:Hfactor}, to compute the values of $H(t_j)$ and $H(t_k)$ for the right-hand side of Eq.~\ref{eq:iteratemethod}.
    \item Using the computed $H$ values, we solve for the left-hand side, $\frac{\overline{\svlu}_{t=0}}{\ig}$, of Eq.~\ref{eq:iteratemethod}.
    \item We then fit the computed results, $\frac{\overline{\svlu}_{t=0}}{\ig}$, in step 2 to the universal fitting function to obtain fitted values $\svt$ and $\ig$.
    \item We replace the initial guessed value for $\svt$ with the newly fitted result and repeat the steps from 2 to 4 until the fitted $\svt$ converges.
\end{enumerate}
We have found this iteration method works well and observed a convergence of the $\svt$ value after 3 steps even when the initial guess was off by a factor of 6, as illustrated by the simulated results shown in figure~\ref{fig:iterationmethod}.  Convergence happens because if the initial guess of $\svt$ is overestimated (underestimated), the heating rate will also be overestimated (underestimated) which leads to a smaller (larger) fitted $\svt$ result.  Therefore, the guessed value at each iteration step will be closer to the true value.  Using this method we could determine the $\svt$ value with systematic errors from heating removed.

\section{Quantum Scattering Calculations}

To describe the quantum scattering dynamics for  {Li+\ce{H2}} and  {Rb+\ce{H2}} collisions, we solve the time-independent Schr\"{o}dinger equation using the coupled channel approach of Arthurs and Dalgarno \cite{arthurs_dalgarno}.

 We represent the interaction potential energy of both  {Li+\ce{H2}} and  {Rb+\ce{H2}} as a Legendre polynomial expansion,
 \begin{equation}\label{eqn:exp}
     U(R,\theta) = \sum_{\lambda} U_{\lambda}(R) P_{\lambda}(\mathrm{cos}\theta),
 \end{equation}
% %
where $R$ is the magnitude of the centre of mass (COM) of \ce{H2} to the COM of the atom, and $\theta$ is the angle between $\vec{R}$ and the molecular axis.
For  {Li+\ce{H2}}, the coefficients $U_{\lambda}(R)$ are provided by Makrides \textit{et al.} for $\lambda=0,2,4$ and $R<20$ a.u.\cite{PhysRevA.99.042704, PhysRevA.105.039903} 
%To interpolate and extrapolate these functions to 50 a.u., we fit the long-range points to a Degli-Esposti-Werner function: 
% \begin{equation}
% \begin{split}
%      {f(x)} & {= (g_0+g_1x+g_2x^2+g_3x^3+g_4x^4+g_5x^5)}\\
%     & {\times \mathrm{exp}(-a_1x-a_2)-\frac{1}{2}(1+\mathrm{tanh}(1+tx))}\\
%     & {\times (C_6x^{-6}+C_8x^{-8}+C_{10}x^{-10}),}
% \end{split}
% \end{equation}
% where in the case of  {Li+\ce{H2}}, the constant $C_6=79.43$ for $\lambda=0$, $C_6=8.912050$ for $\lambda=2$, and $C_6 = 0$ for $\lambda = 4$. In the case of $V_{\lambda=4}(R)$, several short-range data points seemed to inaccurately represent the true potential from $R$=10.5 a.u. to 15 a.u. Therefore, the data from Ref.~\cite{PhysRevA.99.042704} was only used up to 10 a.u. In all cases, the points were interpolated using a spline method.
%
%\begin{figure}[H]
%    \centering
%    \includegraphics[width =\linewidth]{Li_PES.pdf}
%    \caption{ {Legendre expansion coefficients $V_\lambda(R<30~\mathrm{a.u.})$ for (a) $\lambda=0$, (b) $\lambda=2$, and (c) $\lambda=4$.}}
%    \label{?fig:li_PES}
%\end{figure}
%
The  {Rb+\ce{H2}} interaction potential ${U}(R,\theta)$ from Ref.~\cite{Upadhyay2019} is used with permission from Upadhyay \textit{et al.} 
For both systems, we employed a Degli-Esposti-Werner function,
 \begin{equation}
 \begin{split}
      {f(x)} & {= (g_0+g_1x+g_2x^2+g_3x^3+g_4x^4+g_5x^5)}\\
     & {\times \mathrm{exp}(-a_1x-a_2)-\frac{1}{2}(1+\mathrm{tanh}(1+tx))}\\
     & {\times (C_6x^{-6}+C_8x^{-8}+C_{10}x^{-10}),}
 \end{split}
 \end{equation}
in order to interpolate and extrapolate the data \cite{DEW}.

The collision cross sections are computed for energies from 0.1 cm$^{-1}$ to 2000 cm$^{-1}$. 
%The spacing between energies is 0.1 cm$^{-1}$ up until 0.5 cm$^{-1}$, 0.2 cm$^{-1}$ until 9.9 cm$^{-1}$, and 5 cm$^{-1}$ spacing after 10 cm$^{-1}$.  {Energies 0.1 and 0.2~cm$^{-1}$ were excluded from the Rb-\ce{H2} calculations, which does not affect the final calculated values within the error specified.}
Calculations for a given initial rotational state $N_i$ were performed seperately, for $N_i = 0 - 6$. 
% This selection of initial states was determined by plotting the Boltzmann factor as a function of $N_i$,
% \textcolor{black}{
% \begin{equation}
% \begin{split}
%     f_{B}(E_{\mathrm{rot}}) & = \mathrm{exp}(-\beta \times E_{\mathrm{rot}})\\
%     E_{\mathrm{rot}} & = B_e N_i(N_i+1),
% \end{split}
% \end{equation}}
% where $\beta=1/(k_BT)$ with $T=300K$ and the Boltzmann constant $k_B$. We chose $N_{\mathrm{max}}=6$ since this is a couple $N$ levels higher than where the Boltzmann factor is effectively zero. 
Additionally, to account for the two spin isomers of molecular Hydrogen, calculations with orthohydrogen can only have odd values of the $N$ (1,3,...) and parahydrogen can only have even values of $N$ (0,2,...). This is because protons are fermions which require an antisymmetric wavefunction; since orthohydrogen has symmetric nuclear spin, it must have antisymmetric rotational wavefunctions, and since parahydrogen has antisymmetric nuclear spin, it must have symmetric rotational wavefunctions. 

For  {Li+\ce{H2}}, a basis of $J=0 - 125$ is employed. A calculation with a given initial state $N_i$, employs a basis with the maximum rotational quantum number $N_{\mathrm{max}}$=12. For the $R$-propagaton procedure, we employ a grid from 2 a.u. to 50 a.u. with 0.0025 a.u. spacing.

 {For Rb+\ce{H2}, we use a basis with $J_{\mathrm{max}}=300$ and $N_{\mathrm{max}}$=12. The $R$-propagation grid goes from 4 a.u. to 120 a.u. for collision energies greater than 1~cm$^{-1}$, and 4 a.u. to 250 a.u. for energies less than 1~cm$^{-1}$.}

\begin{figure}[H]
    \centering
    \includegraphics[width=\linewidth]{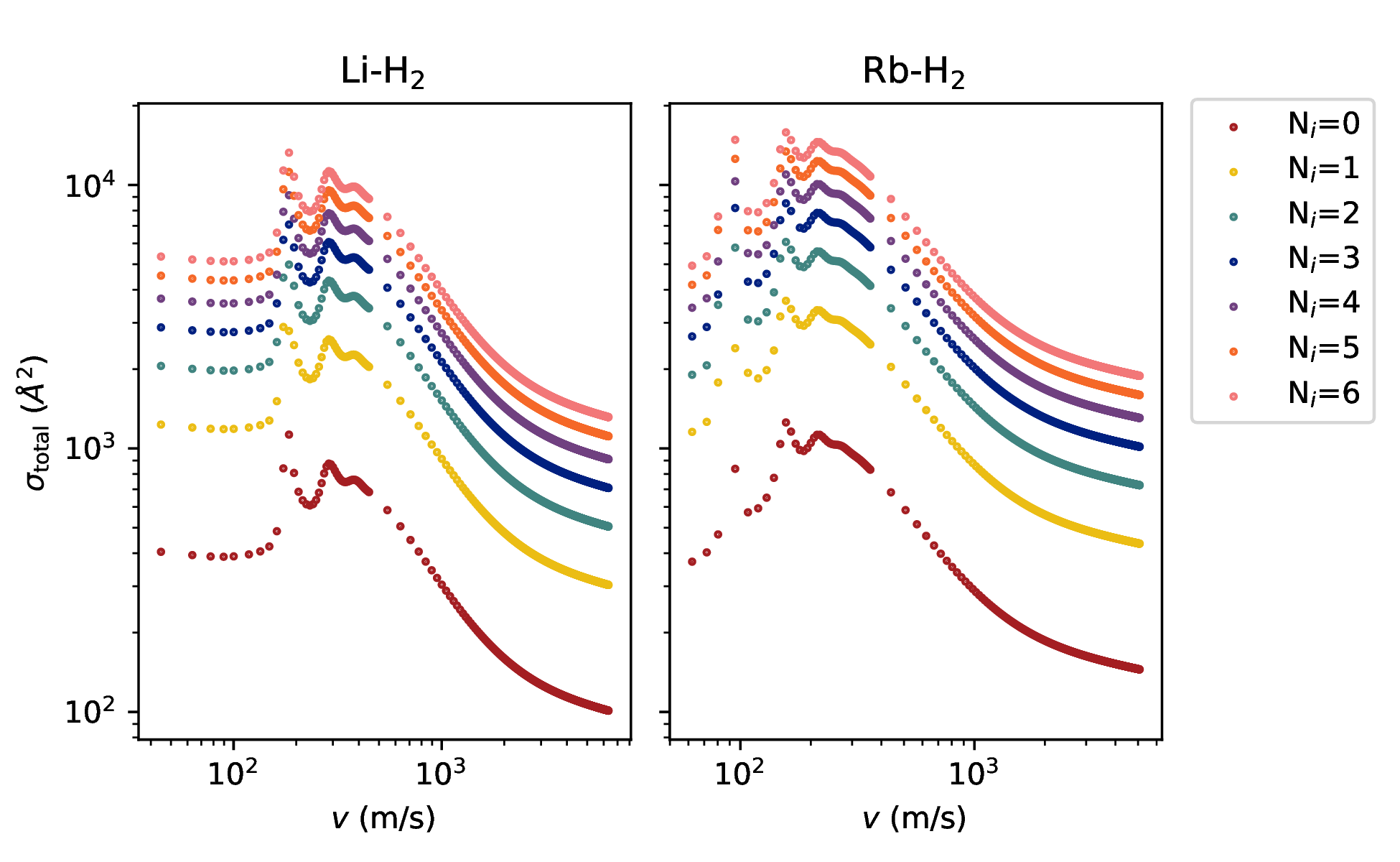}
    \caption{Total cross-section in log-log format plotted as a function of velocity for Li+\ce{H2} (left) and Rb+\ce{H2} (right). The legend differentiates the different initial rotational states $N_i$.}
    \label{fig:theory_fig_2}
\end{figure}

\begin{figure}[H]
    \centering
    \includegraphics[width=\linewidth]{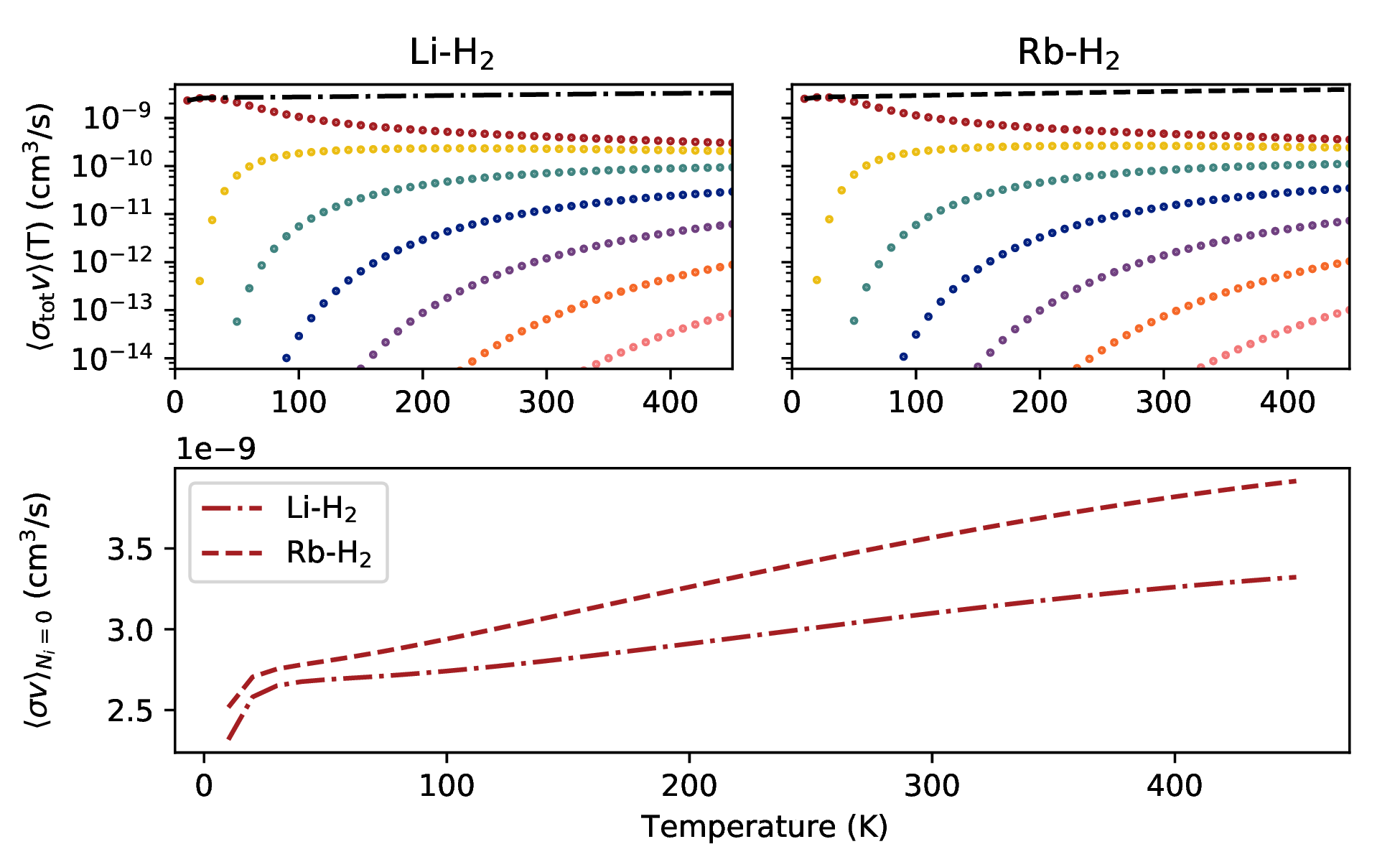}
    \caption{Top row: the averaged thermalized rate coefficient is plotted as a function of temperature for Li+\ce{H2} (top left) and Rb+\ce{H2} (top right). The final theory values that we quote in the paper are taken from the thick black line at 300~K in these subfigures. Bottom row: Equation \ref{L_eqn} plotted for initial state $N_i=0$ only, as the curve for the other initial states are almost overlapping.}
    \label{fig:theory_fig_1}
\end{figure}

Given the total cross-section $\sigma_{N_i}(E_{\mathrm{col}})$ from the scattering calculations (see Figure \ref{fig:theory_fig_2}), our goal is to determine the averaged thermalized rate coefficient
\begin{equation}
    K_{\mathrm{all}}(T) = \frac{\Gamma}{n_{\mathrm{BG}}},
\end{equation}
where $\Gamma$ is the trap loss rate and $n_{\mathrm{BG}}$ is the background gas density. We first calculate the total collision rate coefficient $\langle \sigma v \rangle_{N_i} $ for a given initial state by integrating the product of the cross-section and the velocity $v$ of the molecule over the Maxwell-Boltzmann distribution for collision energy $p(E_{\mathrm{col}})$,
\begin{equation}
    \langle \sigma v \rangle_{N_i} \;=  \int \sigma(E_{\mathrm{col}}) v_{N_i} p(E_{\mathrm{col}}) dE_{\mathrm{col}}
\end{equation}
where
\begin{equation}
    p(E_{\mathrm{col}}) = 2\sqrt{\frac{E_{\mathrm{col}}}{\pi}}\left (\frac{1}{k_BT} \right )^{3/2}\mathrm{exp}\left (-\frac{E_{\mathrm{col}}}{k_BT}\right).
\end{equation}
Next, we let $L_{N_i}(T)$ (see Figure \ref{fig:theory_fig_1}) denote the thermally-averaged rate coefficient for a given initial state,
\begin{equation}\label{L_eqn}
\begin{split}
    L_{N_i}(T) = \frac{\mathrm{exp}(-E_{N_i}/k_BT)}{Z} \langle \sigma v \rangle_{N_i},
\end{split}
\end{equation}
which includes the partition function
\begin{equation}
    Z = \sum_{N}(2N+1) (2-(-1)^N)\mathrm{exp}(-E_{N}/k_BT)
\end{equation}
for all possible initial states $N=0 \rightarrow 6$. In the above equation, $(2-(-1)^N)$ accounts for the ratio 3:1 of orthohydrogen to parahydrogen. Then, the averaged thermalized rate coefficient is
\begin{equation}
    %K_{\mathrm{all}}(T) 
    \svt = \sum_{N_i}(2N_i+1)(2-(-1)^{N_i})\times L_{N_i}(T).
\end{equation}

 {For room temperature $T=300$~K, we calculated $\svt \approx 3.570  \times 10^{-15} \; \mathrm{m}^3/\mathrm{s}$ for Rb+\ce{H2} %3.569909750237534
and $\svt \approx 3.101 \times10^{-15}$ m$^3$/s for Li+\ce{H2}, converged to four significant figures.  %3.100758345330501
Then, the ratio 
\begin{equation}
\begin{split}
    R & = \frac{\svt ^{\mathrm{Li}+\ce{H2}}(T=300~\mathrm{K})}{\svt ^{\mathrm{Rb}+\ce{H2}}(T=300~\mathrm{K})} \\
    & = \frac{3.101\times10^{-15} \;\mathrm{m}^3/\mathrm{s}}{3.570\times10^{-15}\;\mathrm{m}^3/\mathrm{s}}\\
    & = 0.8686
\end{split}
\end{equation}
is within the error bars of the experiment. }

%\begin{figure}[H]
%    \centering
%    \includegraphics[width=\linewidth]{TotalCS_degen.pdf}
%    \caption{Li-\ce{H2}: total cross-section as a function of collision energy for various initial states.}
%\end{figure}
%\begin{figure}[H]
%    \centering
%    \includegraphics[width=\linewidth]{averaging.pdf}
%    \caption{Li-\ce{H2}: total collision rate coefficient as a function of temperature  for various initial states.}
%\end{figure}
%\begin{figure}[H]
%    \centering
%    \includegraphics[width=\linewidth]{KCAVS_new2.pdf}
%    \caption{Li-\ce{H2}: state-averaged thermalized rate coefficient (thick line) and thermally-averaged rate coefficients for all initial states (scatter plots) as a function of temperature for various initial states.}
%\end{figure}


%\bibliography{RbLiComparison}

\end{document}